\documentclass[12pt]{article}
\usepackage{graphicx}
\usepackage{epstopdf}
\usepackage{ccaption}
\usepackage{a4wide}
\usepackage[centertags]{amsmath}
\usepackage{amssymb}
\usepackage[sort&compress,numbers]{natbib}
\usepackage{ifpdf}
\usepackage{setspace}
\usepackage{amsfonts}
\usepackage{color}
\usepackage{threeparttable}
\ifpdf
\usepackage[colorlinks=true,
      linkcolor=black,
      citecolor=blue,
      urlcolor=blue,
      filecolor=blue,
      pdfstartview=FitV,
      pdftitle={},
        pdfauthor={},
        pdfsubject={Hydrodynamics},
        pdfkeywords={hydrodynamics, entropy current, AdS/CFT},
        pdfpagemode=None,
        bookmarksopen=true
      ]{hyperref}
\else
\usepackage[hypertex]{hyperref}  
\fi
\usepackage{graphicx}

\makeatletter
\@addtoreset{equation}{section}

\makeatletter
\renewcommand\section{\@startsection {section}{1}{\z@}%
                                   {-3.5ex \@plus -1ex \@minus -.2ex}
                                   {2.3ex \@plus.2ex}%
                                   {\normalfont\large\bfseries}}
\renewcommand\subsection{\@startsection{subsection}{2}{\z@}%
                                     {-3.25ex\@plus -1ex \@minus -.2ex}%
                                     {1.5ex \@plus .2ex}%
                                     {\normalfont\bfseries}}

  \captionnamefont{\bfseries}
  \captiontitlefont{\small\sffamily}
  \captiondelim{: }
  \hangcaption

 \newcommand{\CO}{\mathcal{O}}

 \newcommand{\CD}{\mathcal{D}}

\newcommand{\bs}{{\bar{\sigma}}}

\def\sec#1{\S \;\ref{#1}}
\def\fig#1{Fig.\,\ref{#1}}
\def\req#1{(\ref{#1})}
\def\App#1{Appendix \ref{#1}}

\def\ie{{\it i.e.}}
\def\etc{{\it etc.}}
\def\eg{{\it e.g.}}
\def\l{\ell}
\def\cf{{\it cf}}
\def\viz{{\it viz.}}
\def\CS{{\mathcal S}}
\def\CO{{\mathcal O}}
\def\CH{{\mathcal H}}

\def\bs{{\bf s}}
\def\bv{{\bf v}}

\def\bCV{{\mathfrak V}}
\def\bCS{{\mathfrak S}}

\def\Ss#1{\bs{\mathnormal #1}}
\def\ST#1{\bCS{\mathnormal #1}}
\def\V#1{\bv{\mathnormal #1}}
\def\VT#1{\bCV{\mathnormal #1}}

\title{Local Fluid Dynamical Entropy from Gravity}

\author{Sayantani Bhattacharyya$^a$, \ Veronika E Hubeny$^b$,
R. Loganayagam$^a$ \\[1.5mm]
Gautam Mandal$^a$, \
Shiraz Minwalla$^a$,\ 
Takeshi Morita$^a$,\\[1.5mm]
Mukund Rangamani$^b$, \
and Harvey S. Reall$^c$\footnote{{\tt   sayanta@theory.tifr.res.in, veronika.hubeny@durham.ac.uk, nayagam@theory.tifr.res.in,
\newline mandal@theory.tifr.res.in, minwalla@theory.tifr.res.in, takeshi@theory.tifr.res.in,\newline mukund.rangamani@durham.ac.uk, hsr1000@cam.ac.uk}}
\\ \\ \\
\small{\emph{$^{a}$ Department of Theoretical Physics,Tata Institute of Fundamental Research,}} \\
\small{\emph{Homi Bhabha Rd, Mumbai 400005, India}} \\ [1mm]
\small{\emph{$^{b}$ Centre for Particle Theory \& Department of
Mathematical Sciences}}, \\
\small{\emph{Science Laboratories, South Road, Durham DH1 3LE, United
    Kingdom}} \\[1mm]
\small{\emph{$^{c}$ Department of Applied Mathematics and Theoretical
    Physics, Centre for Mathematical Sciences,}} \\ 
\small{\emph{Wilberforce Road, Cambridge CB3 0WA, United Kingdom}}
}
\begin{document}

\setlength{\baselineskip}{16pt}
\begin{titlepage}
\maketitle
\begin{picture}(0,0)(0,0)
\put(350,375){TIFR/TH/08-06}
\put(350, 360){DCPT-08/15}
\end{picture}
\vspace{-36pt}


\begin{abstract}

Spacetime geometries dual to arbitrary fluid flows in  strongly
coupled ${\cal N}=4$ super Yang Mills theory  have recently been
constructed perturbatively in the long wavelength limit. We
demonstrate that these geometries all have regular event  horizons,
and determine the location of the horizon order by order in a boundary
 derivative expansion. Intriguingly,  the derivative expansion allows
us to determine the location of the event horizon in the bulk as a
local function of the fluid dynamical variables. We define a natural
map from the boundary to the horizon using ingoing null geodesics. The
area-form on spatial sections of the horizon can then be pulled back
to the boundary to define a local entropy current for the dual field
theory in the hydrodynamic limit. The  area theorem of general
relativity guarantees the positivity of the divergence of the  entropy
current thus constructed.
\end{abstract}

\thispagestyle{empty}
\setcounter{page}{0}
\end{titlepage}

\renewcommand{\baselinestretch}{1}  
\tableofcontents
\renewcommand{\baselinestretch}{1.2}  

\section{Introduction}
\label{intro}

Over the last few years, a special class of strongly coupled $d$-dimensional 
conformal field theories have been ``solved'' via the AdS/CFT duality. Quite 
remarkably, the solution to these theories is given by the equations of $d+1$ 
dimensional gravity (interacting with other fields) in  AdS$_{d+1}$ spacetime. 
Since the long distance dynamics of any genuinely interacting field theory is well described by the equations of relativistic hydrodynamics, it follows as a prediction of the AdS/CFT correspondence that at long distances, the equations of gravity in an AdS$_{d+1}$ background should reduce to the (relativistic) Navier-Stokes equations in $d$ dimensions. There is now substantial direct evidence for the connection between the long distance equations of gravity on AdS$_{d+1}$ spacetime and $d$ dimensional relativistic fluid dynamics; \cf., \cite{Policastro:2001yc,Son:2002sd,Policastro:2002se,Policastro:2002tn,Herzog:2002pc,Herzog:2002fn,Herzog:2003ke,Kovtun:2003wp,Buchel:2003tz,Buchel:2004di,Buchel:2004qq,Kovtun:2004de,Kovtun:2005ev,Benincasa:2005iv,Maeda:2006by,Mas:2006dy,Saremi:2006ep,Son:2006em,Benincasa:2006fu, Janik:2005zt,Janik:2006gp, Nakamura:2006ih, Bhattacharyya:2007vs, Sin:2006pv, Janik:2006ft, Friess:2006kw, Kajantie:2006ya, Heller:2007qt,Kajantie:2007bn,Chesler:2007sv,Benincasa:2007tp,Baier:2007ix,Bhattacharyya:2007jc, Natsuume:2007ty,Natsuume:2008iy,Buchel:2008ac,VanRaamsdonk:2008fp, Gubser:2008vz} for a sampling of the literature on the subject.

In particular, it was noted in \cite{Bhattacharyya:2007jc} that the equations 
of pure gravity with a negative cosmological constant form a universal subsector 
in any theory of gravity on AdS spacetime. Following up on earlier work \cite{Janik:2005zt,Janik:2006gp, Bhattacharyya:2007vs}, it was demonstrated in \cite{Bhattacharyya:2007jc} (for AdS$_5$) and 
more recently in \cite{VanRaamsdonk:2008fp} (for AdS$_4$) that Einstein's equations in this universal sector may be recast, order by order in a boundary derivative expansion, into equations of motion for two collective fields, namely -- the `temperature' and the `velocity'. These new equations of motion turn out to be simply the relativistic Navier-Stokes equations of fluid dynamics.

The gravitational solutions of \cite{Bhattacharyya:2007jc} and 
\cite{VanRaamsdonk:2008fp} constitute an explicit map from the 
space of solutions of the hydrodynamic equations to the space of long 
wavelength gravitational solutions (which are asymptotically AdS).\footnote{By `long wavelength' solutions, we mean  solutions whose spacetime 
variations are slow on a scale set by their respective boundary extrinsic 
curvature. Via the AdS/CFT dictionary this is the same as the requirement 
that the solutions vary slowly on the scale of the inverse temperature 
associated with the local energy density of the solution.} Subject to a 
regularity condition that we will discuss further below, the solutions 
of \cite{Bhattacharyya:2007jc, VanRaamsdonk:2008fp} are locally exhaustive 
in solution space \ie, all long wavelength solutions to Einstein's 
equations that lie nearby in solution space to a metric dual to a particular 
fluid flow are themselves metrics dual to slightly perturbed fluid flows. 
This at first sight surprising result is a consequence of the requirement 
of regularity. This requirement cuts down the 9-parameter space of 
Fefferman-Graham type solutions of AdS$_5$ spacetime -- parameterized by a 
traceless boundary stress tensor -- to the 4-parameter set of solutions of 
fluid dynamics. 

We believe the local exhaustiveness of the gravity solutions dual to fluid 
dynamics, described in the previous paragraph, in fact generalizes to 
a global statement. We think it likely, in other words, that
 the solutions of \cite{Bhattacharyya:2007jc, 
VanRaamsdonk:2008fp} in fact constitute all long wavelength asymptotically 
AdS solutions of gravity with a cosmological constant; we pause here 
to explain why. Every state in a conformal field theory has an associated 
local energy density and a consequent associated mean free path length scale $l_{mfp}$, the inverse of the temperature that is thermodynamically associated with 
this energy density. As a consequence of interactions every state is expected 
to evolve rapidly -- on the time scale $l_{mfp}$ -- towards local 
thermodynamical equilibrium, in an 
appropriate coarse grained sense,\footnote{The $N \to \infty$
limit of the field theory (dual to the classical limit in gravity) justifies 
this coarse graining and supresses consequent fluctuations (which are dual 
to quantum fluctuations in gravity).} at the local value 
of the temperature. This approach to local equilibrium is not long 
wavelength in time and is not well described by fluid dynamics. The dual 
bulk description of this (short wavelength) phenomenon is presumably 
gravitational collapse into a black hole. On the other hand, once 
local equilibrium has been achieved (\ie, a black hole has been formed) 
the system (if un-forced) slowly relaxes towards global equilibrium. 
This relaxation process happens on length 
and time scales that are both large compared to the inverse local temperature, 
and is well described by fluid dynamics and therefore by the solutions of 
\cite{Bhattacharyya:2007jc, VanRaamsdonk:2008fp}. In other words it seems 
plausible that {\it all} field theory evolutions that are long wavelength 
in time as well as space are locally equilibriated, and so are well described
by fluid dynamics. The discussions of this 
paragraph, coupled with the AdS/CFT correspondence, motivate the conjecture 
that the solutions of \cite{Bhattacharyya:2007jc, VanRaamsdonk:2008fp} are 
the most general regular long wavelength solutions to Einstein's equations 
in a spacetime with negative cosmological constant in five and four 
spacetime dimensions respectively.

\begin{figure}[h!]
 \begin{center}
 \includegraphics[scale=0.75]{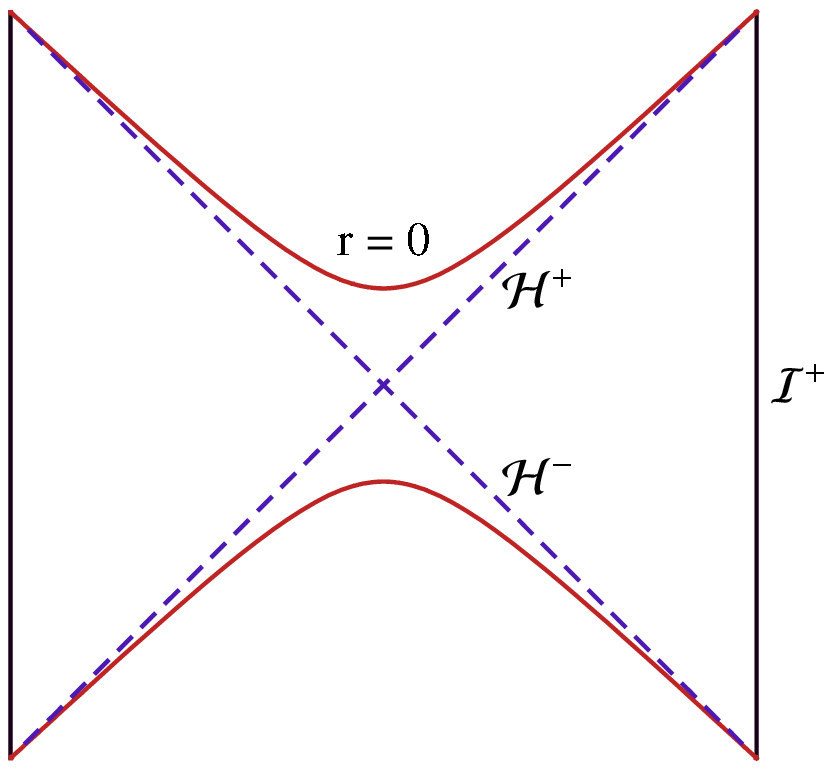}
\hspace{1.5cm}
 \includegraphics[scale=0.33]{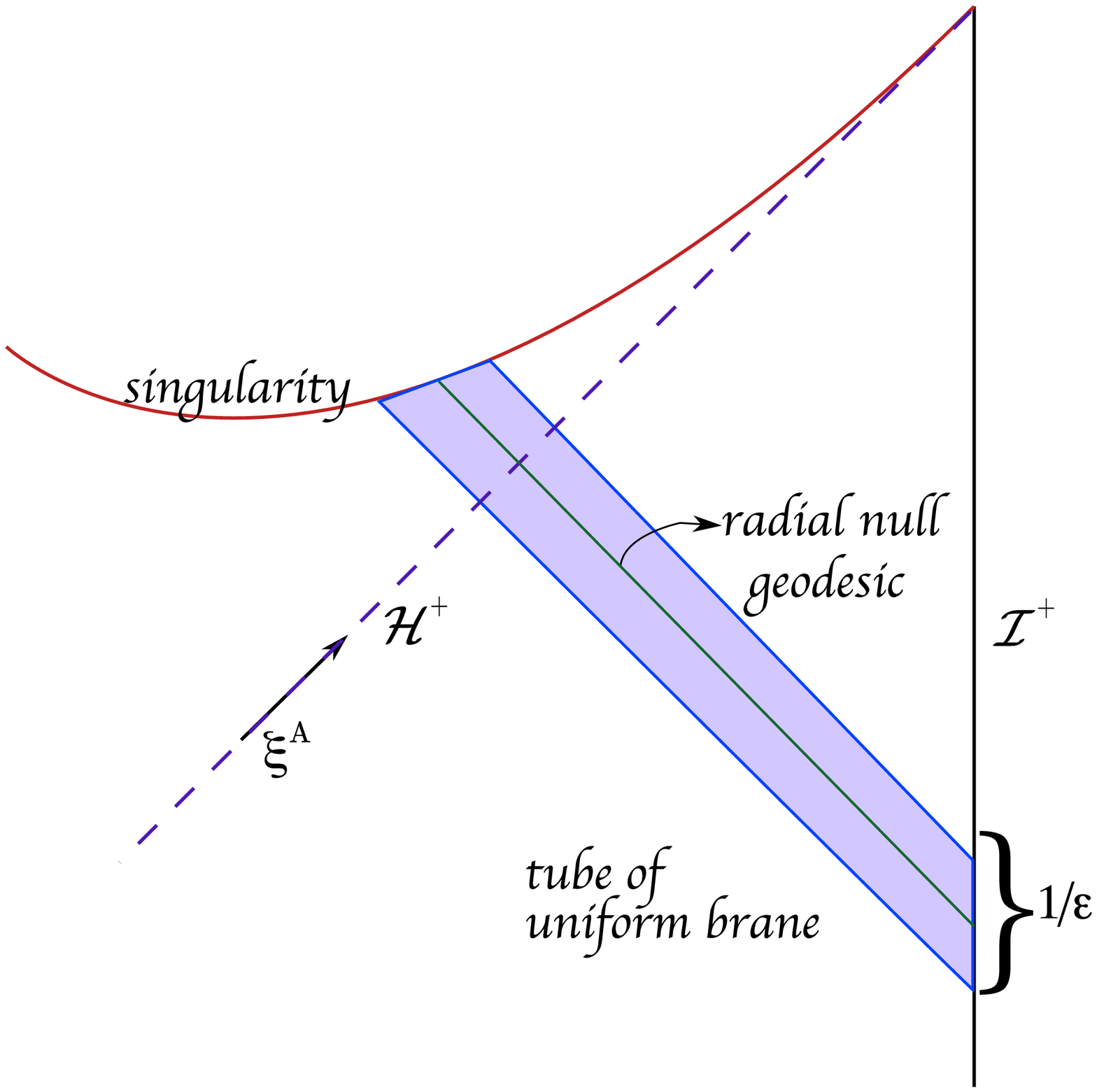} 
\end{center}
\caption{Penrose diagram of the uniform black brane and the causal structure of the spacetimes dual to fluid mechanics illustrating the tube structure. The dashed line in the second figure denotes the future event horizon, while the shaded tube indicates the region of spacetime over which the solution is well approximated by a tube of the uniform black brane.}
\label{PDtube}
\end{figure}

We pause here to note two aspects of the solutions of 
\cite{Bhattacharyya:2007jc,  VanRaamsdonk:2008fp} that we will have 
occasion to use below. First, it is possible to foliate these solutions into 
a collection of tubes, each of which is centered about a radial ingoing null 
geodesic emanating from the AdS boundary. This is sketched in \fig{PDtube} for 
a uniform black brane, where we indicate the tubes on a local portion of the 
spacetime Penrose diagram.\footnote{This is a causal diagram  which captures the entire globally extended spacetime (note that in order for the boundaries to be drawn 
straight, the singularities are curved, as discussed in 
\cite{Fidkowski:2003nf}). For a realistic collapse scenario, described 
by the nonuniform solutions of \cite{Bhattacharyya:2007jc}, only the 
right asymptotic region and  the future horizon and singularity are 
present.} As we will explain below, the congruence of null 
geodesics (around which each of our tubes is centered) yields a natural map 
from the boundary of AdS space to the horizon of our solutions. 
When the width of these tubes in the boundary directions  is small relative 
to the scale of variation of the dual hydrodynamic configurations, the 
restriction of the solution to any one tube is well-approximated tube-wise 
by the metric of a uniform brane with the local value of temperature and 
velocity. This feature of the solutions  -- the fact that they are  
tube-wise indistinguishable from uniform black brane solutions -- is dual 
to the fact that the Navier-Stokes equations describe the dynamics of 
locally equilibrated lumps of fluid. 

Second, the gravitational solutions constructed in \cite{Bhattacharyya:2007jc} are regular everywhere away from a spacelike surface, and moreover the authors conjectured that this singularity is shielded from the boundary of AdS space by a regular event horizon. We will prove this conjecture by explicitly constructing the event horizon of the solutions of \cite{Bhattacharyya:2007jc} order by order in the derivative expansion. It should be possible to carry out a parallel study for the solutions presented in \cite{VanRaamsdonk:2008fp} for four dimensions.  We will not carry out such a study here; however, aspects of our discussion are not specific to AdS$_5$ and can be used to infer the desired features of $2+1$ dimensional hydrodynamics. We expect that the results of such a study would be similar to those presented in this paper. 

As we have explained above,  we study the causal properties -- in particular, the structure of the event horizon for the solutions presented in \cite{Bhattacharyya:2007jc}. We then proceed to investigate various aspects of the dynamics -- specifically, the entropy production -- at this event horizon. In the rest of the introduction, we will describe the contents of this paper in some detail, summarizing the salient points.

As we have discussed above, \cite{Bhattacharyya:2007jc} provides a map 
from the space of solutions of fluid dynamics to a spacetime that solves       
Einstein's equations. The geometry we obtain out of this map depends on 
the specific solution of fluid dynamics we input. In this paper we 
restrict attention to fluid dynamical configurations that approach uniform 
homogeneous flow at some fixed velocity $u_\mu^{(0)}$ and temperature $T^{(0)}$ 
at spatial infinity. It seems intuitively clear from the dissipative 
nature of the Navier-Stokes equations that the late time behaviour of all 
fluid flows with these boundary conditions will eventually become 
$u_\mu(x)=u_\mu^{(0)}$ and $T(x)=T^{(0)}$; we assume this in what 
follows.\footnote{This is true for conformal fluid dynamics in 
$d$ spacetime dimensions for $d \geq 3$. Conformal fluid dynamics in 
$1+1$ dimensions is not dissipative (for instance it is non-viscous since here the shear tensor does not exist). More generally, there 
are as many degrees of freedom in a traceless $1+1$ dimensional stress 
tensor as in temperature and velocity fields in $1+1$ dimensions. 
Consequently, the most general solution to $1+1$ dimensional `fluid 
dynamics' is simply given by $T_{++}=f(\sigma^+)$ and $T_{--}= 
h(\sigma^-)$ for arbitrary functions $g$ and $h$. This solution describes 
left and right moving waves that maintain their shape forever, propagating non-dissipatively  at the speed of light. We thank A. 
Strominger discussions on this point.} The gravitational dual to such an 
equilibrated fluid flow is simply the metric of a uniformly boosted black 
brane.

The causal structure of the uniform black brane is given by the Penrose 
diagram plotted in \fig{PDtube} (see \cite{Fidkowski:2003nf}). In 
particular, the equation for the event horizon of a uniform black brane 
is well known.  The event horizon of the metric dual to the full fluid flow 
is simply the unique null hypersurface that joins with this late time 
event horizon in the asymptotic future.\footnote{Note that for a generic 
hydrodynamic solution, the bulk spacetime has no manifest isometries; the 
event horizon is therefore not a Killing horizon.} It turns out to be 
surprisingly simple to construct this hypersurface order by order 
in the boundary derivative expansion used in \cite{Bhattacharyya:2007jc}. 
In this paper, we perform this construction up to second order in derivatives. 
Within the derivative expansion it turns out that the radial location  of the 
event horizon is determined locally by values and derivatives of fluid dynamical velocity and temperature at the corresponding boundary point. This is achieved using the  boundary to horizon map generated by the congruence of ingoing null geodesics described above (see \fig{PDtube}).\footnote{This 
map may be motivated as follows. Consider perturbing the fluid at a boundary 
point $x^\mu$, \eg, by turning on some local operator of Yang Mills theory. This perturbation instanteneously alters  all fluid quantities, including the entropy, at $x^\mu$. However, it only alters the geometry near 
the horizon at and  within the lightcone emanating from $x^\mu$ at the boundary.   
It is therefore plausible that local properties of the spacetime in the neighbourhood
of ingoing geodesics that emanate from $x^\mu$ capture properties of the 
fluid at $x^\mu$. } 

However, while locality is manifest in a derivative expansion, upon summing all orders
 we expect this local behaviour to  transmute into a limited nonlocality: the radial position of the event  horizon at a given point should be determined by the values of fluid dynamical  variables in a patch of size $1/T$ centered around the associated boundary point. The ateleological behaviour of the event horizon is a surprising feature of these solutions and implies that the  event horizon behaves as a `membrane' whose vibrations are a local mirror of fluid  dynamics. 
Our explicit construction of the event horizon of the metrics dual to fluid 
dynamics is one of the key results of our paper; \cf., \req{simanto}.

We now turn to a description of our second main result; the construction of an entropy current with non-negative divergence for a class of asymptotically AdS solutions of gravity, and its explicit evaluation  for the solutions of \cite{Bhattacharyya:2007jc} at second order in the derivative expansion.\footnote{We have been informed that A. Strominger and S. Hartnoll have independently discussed the construction of a positive divergence entropy  current on the horizon, utilizing Brown York stress tensor.} As we will see in \sec{sec:locecur}, it is possible to define a
natural area  $(d-1)$-form on any event horizon in a $d+1$ dimensional
spacetime in general relativity. This form is defined to ensure that
its integral over any co-dimension one spatial slice of the horizon is
simply the  area of that submanifold. It follows almost immediately
from the definition  of this form and the classic area increase
theorems of general relativity  that the exterior derivative (on the
event horizon) of this $(d-1)$-form, viewed of as a top  dimensional
form on  the horizon, is `positive' (we explain what we mean by the
positivity of  a top form on the horizon in \sec{sec:areaf}).

The positivity of the exterior derivative of the area $(d-1)$-form is a formally elegant restatement of the area increase theorem of general relativity that is local on the horizon. Hence we would like to link this statement to the positivity of the entropy production in the boundary theory. However, at least naively, the CFT fluid dual to our solutions lives at the boundary of AdS space rather than on its horizon. If we wish to study the interplay between the local notion of entropy of the fluid and the fluid equations of motion, it is important for these quantities to be defined on the same space. In order to achieve this, in \sec{sec:bdymap} we use a congruence of null geodesics described above to 
provide a `natural' map from the boundary to the horizon for a class of asymptotically AdS solutions of gravity (which include but are slightly more general than those of \cite{Bhattacharyya:2007jc}). The pullback of the area $(d-1)$-form under this map now lives at the boundary, and also has a `positive' exterior derivative. Consequently, the `entropy current',  defined as the boundary Hodge dual to the pull-back of the area $(d-1)$-form on the boundary (with appropriate factors of Newton's constant), has non-negative divergence, and so satisfies a 
crucial physical requirement for an entropy current of fluid dynamics. 

In \sec{sec:fluid}, we then proceed to implement the construction described in the previous paragraph for the solutions of \cite{Bhattacharyya:2007jc}. This enables us to derive an  expression for the entropy current, $J^\mu_S$, with non-negative divergence, valid up to second order in the derivative expansion. As a check of our final result, we use the equations of fluid dynamics
to independently verify the non-negativity of divergence of our entropy current at third order in the derivative expansion. An example of an entropy current for a conformal fluid with non-negative divergence was first described in \cite{Loganayagam:2008is}.

We also take the opportunity to extend and complete the analysis presented in \cite{Loganayagam:2008is} to find the most general Weyl covariant two derivative
 entropy current consistent with the second law. Note that the requirement 
of pointwise  non-negativity of the  entropy production -- which we impose as 
a physical constraint of acceptable entropy currents -- 
 carries useful information even within the derivative expansion, though 
this is a little subtle to unravel. In particular, in \sec{sec:positive} we present a parameterization of the most general (7 parameter) class of Weyl invariant candidate entropy currents that has the correct equilibrium limit, to second order in the derivative expansion. We also demonstrate that only a five dimensional sub-class of these currents is consistent with the requirement of pointwise 
non-negativity of $\partial_\mu J_S^\mu$ to third order in derivatives. We then turn our attention to the arbitrariness of our gravitational construction of the entropy current and demonstrate that there appears to be a two  parameter family of  physically acceptable generalizations of this bulk construction (associated with physically acceptable generalizations of the boundary to horizon map and also the generalisations of the area $(d-1)$-form itself). As a result, we conclude that the gravitational construction presented in this 
paper yields a two dimensional sub-class in the five dimensional space of entropy currents with non-negative divergence. It would interesting to understand directly from field theory, what principle (if any) underlies the selection of this distinguished class of entropy currents. It would also be interesting to 
investigate whether the remaining positive entropy currents may be obtained 
from a generalized gravitational procedure, perhaps involving apparent and 
other local horizons.\footnote{We thank A. Strominger and M. Van Raamsdonk 
for discussions on this point.}

This paper is organized as follows. In \sec{sec:horizon} below, we present an order by order construction of the event horizon in a class of metrics that include those of \cite{Bhattacharyya:2007jc}. Following that, in \sec{sec:locecur}, we present our construction of a local entropy $(d-1)$-form for the event horizons in gravity, and then implement our  constructions in detail for the class of metrics studied in \sec{sec:horizon}.  In \sec{sec:fluid}, we specialize the results of \sec{sec:horizon} and \sec{sec:locecur} to the metrics dual to fluid dynamics \cite{Bhattacharyya:2007jc}, using a map for translating horizon information to the boundary developed in \sec{sec:bdymap}. We obtain explicit formulae, to second order  in the derivative expansion,  for the event horizon and the entropy current  in the geometries of \cite{Bhattacharyya:2007jc}. In \sec{sec:positive}, we explain in  detail the nature of the constraint imposed on second order terms in the expansion of the entropy current by the requirement of non-negativity of entropy production at third order in the derivative expansion. We also investigate
the relationship between the geometrically constructed entropy current and the general entropy current of non-negative divergence generalizing the analysis of \cite{Loganayagam:2008is}. Finally, in \sec{sec:discuss}, we end with a discussion of  our results and open questions. Some technical results regarding the computations are collected in various Appendices.

\section{The Local Event Horizon}\label{sec:horizon}

As we have explained in the introduction, in this paper we will study the 
event horizon of the metrics dual to fluid dynamics presented in 
\cite{Bhattacharyya:2007jc}. In that reference the authors construct an 
explicit classical spacetime dual to an arbitrary solution of fluid 
dynamics, accurate to second order in the boundary derivative expansion. 
While the explicit solutions of \cite{Bhattacharyya:2007jc} are rather 
involved, we will see below that the structure of the event horizons of 
these solutions are insensitive to many of these details. Consequently, in 
this section we will describe the metric of \cite{Bhattacharyya:2007jc} 
only in general structural form, and carry out all our computations for an 
arbitrary spacetime of this form. In \sec{sec:fluid} we will specialize these 
calculations to the detailed metrics of \cite{Bhattacharyya:2007jc}. We 
start by presenting a geometric interpretation for the 
coordinate system used in \cite{Bhattacharyya:2007jc}.

\subsection{Coordinates adapted to a null geodesic congruence}

Consider a null geodesic congruence (\ie, a family of null geodesics
with exactly one geodesic passing through each point) in some region
of an arbitrary spacetime.
Let $\Sigma$ be a hypersurface
that intersects each geodesic once. Let $x^\mu$ be coordinates on
$\Sigma$. Now ascribe coordinates $(\rho,x^\mu)$ to the point at an 
affine parameter distance $\rho$ from $\Sigma$, along the geodesic through the
point on $\Sigma$ with coordinates $x^\mu$. Hence the geodesics in the
congruence are lines of constant $x^\mu$. In this chart, this metric
takes the form 
\begin{equation}
 ds^2 = -2 \,u_\mu(x) \,d\rho \,dx^\mu +
\widehat{\chi}_{\mu\nu}(\rho,x)\, dx^\mu \,dx^\nu, 
\end{equation}
where the geodesic
equation implies that $u_\mu$ is independent of $\rho$.  It is
convenient to generalize slightly to allow for non-affine
parametrization of the geodesics: let $r$ be a parameter related to
$\rho$ by $d\rho/dr = {\cal S}(r,x)$. Then, in coordinates
$(r,x)$, the metric takes the form\footnote{We use upper case Latin
  indices $\{M,N, \cdots\}$ to denote bulk directions, while lower
  case Greek indices $\{\mu ,\nu, \cdots\}$ will refer to field theory
  or boundary directions. Furthermore, we use lower case Latin indices 
$\{a, b, i,j,\cdots\}$ to denote the spatial directions in the boundary. 
Finally, we use $(x)$ to indicate the dependence on the four coordinates $x^\mu$.
Details regarding the conventions used in this paper can be found in
\App{app:notation}. }  
\begin{equation} 
ds^2 =G_{MN} \,dX^M \,dX^N = - 2 \, u_\mu(x) \, \CS(r,x)\, dr\, dx^\mu + \chi_{\mu \nu}(r,x) \, dx^\mu\, dx^\nu 
\label{formmet}
\end{equation}
Note that $\Sigma$ could be spacelike, timelike, or null. We shall take $\Sigma$
to be timelike. 

This metric has determinant $-{\cal S}^2 \chi^{\mu\nu}
u_\mu u_\nu \det \chi$, where $\chi^{\mu\nu}$ is the inverse of
$\chi_{\mu\nu}$. Hence the metric and its inverse will be smooth if
${\cal S}$, $u_\mu$ and $\chi_{\mu\nu}$ are smooth, with ${\cal S}
\ne 0$, $\chi_{\mu\nu}$ invertible, and $\chi^{\mu \nu} \, u_\mu$ timelike. These conditions are satisfied on, and outside, the
horizons of the solutions that we shall discuss below.

\subsection{Spacetime dual to hydrodynamics}\label{sec:metdesc}

The bulk metric of \cite{Bhattacharyya:2007jc} was obtained in a
coordinate system of the form \eqref{formmet} just described, where the role of
$\Sigma$ is played by the conformal boundary and the null geodesics
are future-directed and ingoing at the boundary. The key assumption
used to derive the solution is that the metric is a slowly varying
function of $x^\mu$ or, more precisely, that
the metric functions have a perturbative expansion (with a small parameter $\epsilon$):
\begin{equation}
\CS(r,x) = 1 - \sum_{k=1}^\infty \, \epsilon^k \, s_a^{(k)}  \ ,
\label{csdefn}
\end{equation}
\begin{equation}
\chi_{\mu \nu}(r,x) = -r^2 \, f(b\,r)\, u_\mu \, u_\nu + r^2 \, P_{\mu \nu} + \sum_{k=1}^\infty \, \epsilon^k \, \left(  s_c^{(k)}\;  r^2 \, P_{\mu\nu} +  s_b^{(k)} \, u_\mu \, u_\nu +\, j^{(k)}_{\nu} \,u_{\mu} + 
j^{(k)}_{\mu} \,u_{\nu}
+ t^{(k)}_{\mu \nu} \right) \ . 
\label{chidef}
\end{equation}	
The function $f(y)$ above has the form $f=1-\frac{1}{y^4}$; however,
the only property of $f$ that we will use is that $f(1)=0$. 
The remaining functions ($s_a^{(k)}$, $s_b^{(k)} \ldots$) are all 
local functions of the inverse temperature $b(x)$ and the velocity $u^\mu(x)$  and 
the coordinate $r$, whose
form was determined in \cite{Bhattacharyya:2007jc} and is indicated below in
\req{listing} and given explicitly in \App{app:notation};  we however will not
need the specific form of these functions for the present discussion. As far as  the calculations in this section are concerned, the expressions  $s_a^{(k)}, s_b^{(k)}, s_c^{(k)}, j^{(k)}_\mu$ and  $t_{\mu\nu}^{(k)}$  may be thought of as arbitrary functions of $r$ and 
$x^\mu$. The tensor $P_{\mu \nu} = \eta_{\mu \nu} + u_{\mu} \, u_\nu$ is a co-moving spatial projector.

In the above formulae, $\epsilon$ is a formal derivative counting parameter. Any expression  that multiplies $\epsilon^k$ in \eqref{csdefn} and \eqref{chidef} is of 
$k^{th}$ order in boundary field theory derivatives. Note that any 
boundary derivative of any of the functions above is always accompanied by an 
additional explicit power of $\epsilon$. As in \cite{Bhattacharyya:2007jc},
all calculations in this paper will be performed order by order in $\epsilon$ which is 
then set to unity in the final results. This is a good approximation when 
field theory derivatives are small in units of the local temperature. 

As we have explained in the Introduction, the metrics presented in \cite{Bhattacharyya:2007jc}  simplify to the uniform black brane metric at late times. This metric describes a fluid configuration with constant $u^\mu$ and $b$. As the 
derivative counting parameter $\epsilon$ vanishes on constant configurations, 
all terms in the summation in \eqref{csdefn} and \eqref{chidef} vanish on the 
uniform black brane configuration. The event horizon of this simplified metric 
is very easy to determine; it is simply the surface $r=\frac{1}{b}$. 
Consequently, the event horizon  $\CH$ of the metric \eqref{formmet} has a 
simple mathematical characterization; it is the unique null hypersurface that 
reduces exactly, at infinite time to $r=\frac{1}{b}$. 
 
In \sec{sec:pertcomp} we will describe a local construction of a null 
hypersurface in the metric \eqref{formmet}. Our hypersurface will have the 
property that it reduces exactly to $r=1/b$ when $u^\mu$ and $b$ are 
constants, and therefore may be identified with the event horizon 
for spacetimes of the form \eqref{formmet} that settle down to constant $u^\mu$ and $b$ at late times, as we expect for metrics dual to fluid dynamics. 
  We will evaluate our result for the metrics of \cite{Bhattacharyya:2007jc} in \sec{sec:fluid} where we will use the explicit expressions for the  functions appearing in \eqref{formmet}.

\subsection{The event horizon in the derivative expansion}
\label{sec:pertcomp}

When $\epsilon$ is set to zero and $b$ and $u_\mu$ are constants, the
surface $r=\frac{1}{b}$ is a null hypersurface in metrics
\eqref{formmet}. We will now determine the corrected equation for this
null hypersurface at small $\epsilon$, order by order in the
$\epsilon$ expansion. As we have explained above, this hypersurface
will be physically interpreted as the event horizon $\CH$ of the
metrics presented in \cite{Bhattacharyya:2007jc}.

The procedure can be illustrated with a simpler
example. Consider the Vaidya spacetime, describing a spherically
symmetric black hole with ingoing null matter:
\begin{equation}
 ds^2 = - \left( 1 - \frac{2\,m(v)}{r} \right) \,dv^2 + 2\, dv \,dr + r^2 \,d\Omega^2 \ .
\end{equation}
Spherical symmetry implies that the horizon is at $r=r(v)$,
with normal $n = dr - \dot{r} \, dv$. Demanding that this be null gives 
$r(v) = 2\, m(v) + 2\, r(v)\,  \dot{r}(v)$, a first order ODE for $r(v)$. Solving this determines the position of the horizon {\it non-locally} in terms of $m(v)$. However, 
if we assume that $m(v)$ is slowly varying and approaches a constant for large $v$, \ie, 
\begin{equation}
\dot{m}(v)={\cal O}(\epsilon)  \ , m \,  \ddot{m} = {\cal O}(\epsilon^2), \; \etc, \qquad {\rm  and} \qquad  \lim_{v \to \infty} m(v) = m_0 
\label{}
\end{equation}	
then we can solve by expanding in derivatives. Consider the ansatz,  $r = 2\, m + a\, m \,\dot{m} + b\, m \,\dot{m}^2 + c\, m^2 \,\ddot{m} + \ldots$, for some constants $a,b,c,\ldots$; it is easy to show that the solution for the horizon is given by $a=8$, $b=64$, $c=32$, \etc. Hence we can obtain a {\it local} expression for the
location of the horizon in a derivative expansion.

Returning to the spacetime of \cite{Bhattacharyya:2007jc}, let us
suppose that the null hypersurface that we are after is given by  the equation
\begin{equation} \label{ehsurf} 
S_{\CH}(r, x) =0 \ , \qquad {\rm with} \qquad 
S_{\CH}(r,x) =r-r_H(x) \ .
\end{equation}
As we are working in a derivative expansion we take
\begin{equation}
r_H(x)=  \frac{1}{b(x)}+ \sum_{k=1}^\infty \, \epsilon^k \,r_{(k)}(x)
\label{rhpert}
\end{equation}	
Let us denote the normal vector to the event horizon by $\xi^A$: by definition,
\begin{equation}
\xi^A = G^{AB} \, \partial_B S_{\CH}(r, x)
\label{nadef}
\end{equation}	
which also has an $\epsilon$ expansion. We will now determine $r_{(k)}(x)$ and $\xi_{(k)}^A(x^\mu)$ order by order in $\epsilon$.
In order to compute the unknown functions $r_{(k)}(x)$ 
we require the normal vector $\xi^A$ to be null, which amounts to simply solving the equation
\begin{equation}\label{logic}
G^{AB}\, \partial_A S_{\CH} \, \partial_B S_{\CH} =0
\end{equation}
order by order in perturbation theory. Note that 
\begin{equation}\label{exps}
dS_\CH = dr - \epsilon\, \partial_\mu r_H \ dx^\mu 
\qquad {\rm where } \qquad
\epsilon\, \partial_\mu r_H = 
-\frac{\epsilon}{b^2}\, \partial_\mu b  + \sum_{n=1}^\infty \epsilon^{n+1}  \,\partial_\mu  r_{(n)} \ .
\end{equation}
In particular, to order $\epsilon^n$,  only the functions $r_{(m)}$ for 
$m\leq n-1$ appear in \eqref{exps}. However, the LHS 
of \eqref{logic} includes a contribution of two factors of $dr$ contracted 
with the metric. This contribution is equal to $G^{rr}$ evaluated at the 
horizon. Expanding this term to order $\epsilon^n$ we find a contribution
$$ \frac{1}{\kappa_1 \, b }\, r_{(n)}$$ where $\kappa_1$ is defined in \eqref{kapidef} below, together with several terms 
that depend on $r_{(m)}$ for $m \leq n-1$. It follows that the expansion of 
\eqref{logic} to $n^{th}$ order in $\epsilon$ yields a simple algebraic 
expression for $r_{(n)}$, in terms of the functions $r_{(1)}, r_{(2)}, \cdots, r_{(n-1)}$  which are determined from lower order computations.

More explicitly, equation \req{logic} gives
$G^{rr} - 2 \, \epsilon\, \partial_\mu r_H \, G^{r \mu} 
+ \epsilon^2 \, \partial_\mu r_H \, \partial_\nu r_H \, G^{\mu \nu} = 0
$, with the inverse metric $G^{MN}$ given by:
\begin{equation}
G^{rr} = \frac{1}{-\CS^2 \,  u_\mu \, u_\nu \, \chi^{\mu\nu}} \ , \qquad
G^{r\alpha} = \frac{\CS \, \chi^{\alpha\beta}\,u_{\beta}}{- \CS^2 \, u_\mu \, u_\nu \, \chi^{\mu\nu}} \ , \qquad
G^{\alpha\beta} = \frac{\CS^2 \, u_ \gamma \, u_ \delta \, \left(
\chi^{\alpha\beta} \,\chi^{\gamma\delta} - \chi^{\alpha \gamma} \,\chi^{\beta \delta}  \right)}{- \CS^2 \, u_\mu \, u_\nu \, \chi^{\mu\nu}} \ .
\label{invmet}
\end{equation}	
where the `inverse $d$-metric' $\chi^{\mu\nu}$ is defined via $\chi_{\mu\nu} \, \chi^{\nu\rho} = \delta_\mu^{\ \rho}$.
Hence the expression for the location of the event horizon  \req{logic} to arbitrary order in $\epsilon$ is obtained by expanding
\begin{equation}
0 =  \frac{1}{-\CS^2 \,  u_\mu \, u_\nu \, \chi^{\mu\nu}} 
\left( 1 - 2 \,\epsilon \, \CS \, \chi^{\alpha\beta}\,u_{\beta} \, \partial_\alpha r_H 
- \epsilon^2 \, \CS^2 \, \left(
\chi^{\alpha\beta} \,\chi^{\gamma\delta} - \chi^{\alpha \gamma} \,\chi^{\beta \delta}  \right) 
\, u_ \gamma \, u_ \delta \, \partial_ \alpha r_H \, \partial_ \beta r_H 
\right)
\label{horizon}
\end{equation}	
to the requisite order in $\epsilon$, using the expansion of the individual  quantities $\CS$ and $r_H$ specified above, as well as of $\chi^{\mu\nu}$.

\subsection{The event horizon at second order in derivatives}
\label{sec:evsecord}

The equation \eqref{logic} is automatically obeyed at order $\epsilon^0$. 
At first order in $\epsilon$ we find that the location of the event horizon
is given by $r=r_H^{(1)}$ with\footnote{We have used here the fact that $u^\mu\,j^{(k)}_\mu  =0$ and $u^\mu\,t^{(k)}_{\mu\nu} = 0$ which follow from the solution of \cite{Bhattacharyya:2007jc}. We also restrict to solutions which are asymptotically AdS$_5$ in this section.} 
\begin{equation}\label{firstorder}
r^{(1)}_H(x)  = \frac{1}{b(x)} + r_{(1)}(x)= 
\frac{1}{b} + \kappa_1\, \left( s_b^{(1)}
-\frac{2}{b^2} \,u^\mu \,\partial_\mu b \right) \ . 
\end{equation}
where we define 
\begin{equation}
\frac{1}{\kappa_m} = \frac{\partial^m}{\partial r^m }\left( r^2\, f(b\,r) \right) \biggr|_{r = \frac{1}{b}}
\label{kapidef}
\end{equation}	
At next order, $\CO(\epsilon^2)$, we find 
\begin{equation}\label{secondorder} \begin{split}
&r^{(2)}_H(x) = \frac{1}{b} + \kappa_1\,\Bigg( s_b^{(1)} + \partial_r s_b^{(1)}  r_H^{(1)} -\frac{2}{b^2}  \, \left(1-  s_a^{(1)} \right) \,u^\mu \,\partial_\mu b + s_b^{(2)} + 2 \, u^\mu\,\partial_\mu
r_{(1)}\\
&\qquad\qquad -\frac{1}{b^2}\, P^{\mu\nu}\,\left(b^2 \,j_{\mu}^{(1)} + \partial_\mu b\right)\,\left(b^2 \,j_{\nu}^{(1)} + \partial_\nu b\right) - \frac{1}{2\, \kappa_2}\,r_{(1)}^2\Bigg)
\end{split}
\end{equation}
where we have\footnote{It is important to note that in our expressions involving the boundary derivatives we raise and lower indices using the boundary metric $\eta_{\mu \nu}$; in particular, $u^\mu \equiv \eta^{\mu \nu}\, u_{\nu}$ and with this defintion $u^\mu \, u_\mu = -1$.} 
 $$ P^{\mu\nu} = u^\mu u^\nu + \eta^{\mu\nu}\qquad  \text{and}\qquad\eta^{\mu\nu} = diag(-1,1,1,1) \ . $$
As  all functions and derivatives in \req{firstorder} and \req{secondorder} are evaluated at  $r=1/b$ and the point $x^\mu$ and we retain terms to $\CO\left(\epsilon\right)$ and $\CO\left(\epsilon^2\right)$ respectively.  

It is now simple in principle to plug \eqref{secondorder} into \eqref{formmet}
to obtain an explicit expression for the metric $H_{\mu \nu}$ of the event horizon.\footnote{There are thus three metrics in play; the bulk metric defined in \eqref{formmet}, the boundary metric which is fixed and chosen to be $\eta_{\mu \nu}$ and finally the metric on the horizon $\CH$, $H_{\mu \nu}$, which we do not explicitly write down. As a result there are differing and often conflicting notions of covariance; we have chosen to write various quantities consistently with boundary covariance since at the end of the day we are interested in the boundary entropy current.}  We will choose to use the coordinates $x^\mu$ to parameterize the event horizon.  The normal vector $\xi^A$ is a vector in the tangent space of the event  horizon (this follows since the hypersurface is null), \ie,
\begin{equation}
\xi^A\, \frac{\partial}{\partial {X^A}}=n^\mu \, \frac{\partial}{\partial{x^\mu}} + n^r\, \frac{\partial}{\partial r} \ ,
\label{nxidef}
\end{equation}	
which is easily obtained by using the definition \eqref{nadef} and the induced metric on the event horizon; namely
\begin{equation}\label{nm1} \begin{split}
n^\mu&= \left( 1+ s_a^{(1)}+ (s_a^{(1)})^2 + s_a^{(2)} \right)u^\mu -\frac{1}{r^4} (t^{(1)})^{\mu\nu} \left( j_\nu^{(1)} +\frac{\partial_\nu b}{b^2} \right) \\
&\qquad  
+  \frac{1}{r^2} P^{\mu \nu} \left( j_\nu^{(1)} \,\left( 1 + s_a^{(1)}
-s_c^{(1)}\right) + \frac{\partial_\nu b}{b^2} \, \left( 1 - s_c^{(1)} \right) +  j_\nu^{(2)} -\partial_\nu r_{(1)} \right) \ .
\end{split}
\end{equation}

Before proceeding to analyze the entropy current associated with the local area-form on this event horizon, let us pause to consider the expression \req{secondorder}.
First of all, we see that for generic fluids with varying temperature and velocity, the radial coordinate $r=r_H$ of the horizon varies with $x^{\mu}$, which, to the first order in the derivative expansion, is given simply by the local temperature.
The constraints on this variation are inherited from the equations of relativistic fluid dynamics which govern the behaviour of these temperature and velocity fields, as discussed above.  Note that the variation of $r_H$ at a given $x^i$ 
and as a function of time, can of course be non-monotonic.  As we will see in the next section, only the local area needs to increase.  This is dual to 
the fact that while a local fluid element may warm up or cool down in 
response to interacting with the neighbouring fluid, the local entropy 
production is always positive. 
\begin{figure}[h!]
\begin{center}
\includegraphics[scale =0.8]{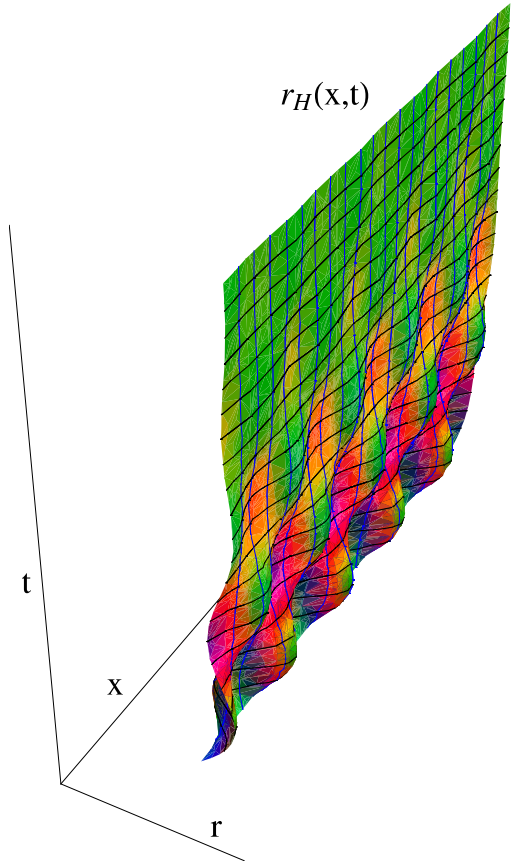}
\end{center}
\caption{The event horizon $r=r_H(x^\mu)$ sketched as a function of the time $t$ and one of the spatial coordinates $x$ (the other two spatial coordinates are suppressed).} 
\label{wigglyhor}
\end{figure}
An example of the behaviour of $r_H(x)$ is sketched in the spacetime diagram of \fig{wigglyhor}, with time plotted vertically and the radial coordinate as well as one of the spatial $x^i$ coordinates plotted horizontally.

\section{The Local Entropy Current}
\label{sec:locecur}

Having determined the location of the event horizon, it is a simple matter to compute the area of 
the event horizon to obtain the area of the black brane. 
However, as we wish to talk about the spatio-temporal variation of the entropy, we will first describe entropy production in a local setting. This will 
allow us to derive an expression for the boundary entropy current in \sec{sec:fluid}.

\subsection{Abstract construction of the area $(d-1)$-form} 
\label{sec:areaf}

In this brief subsection we present the construction of the area $d-1$ 
form on the spatial section of any event horizon of a $d+1$ dimensional solution of general relativity. 

First, recall that the event horizon is a co-dimension one null submanifold of the $d+1$ dimensional spacetime. As a result its normal vector lies in its tangent space. The horizon generators coincide with the integral curves of this normal  vector field, which are in fact null geodesics\footnote{This follows from the fact that the event horizon is the boundary of the past of future infinity $\mathcal{ I}^+$ together with the fact that boundaries of causal sets are generated by null geodesics \cite{Hawking:1973db}.
We pause here to note a technical point regarding the behaviour of the horizon generators:
While by definition these null geodesics generating the event horizon have no future endpoints \cite{Penrose:1967mz}, they do not necessarily remain on the event horizon when extended into the past.  This is because in general dynamical context, these geodesics will have non-zero expansion, and by Raychaudhuri's equation they must therefore caustic in finite affine parameter when extended into the past.  Hence, although the spacetime, and therefore the event horizon, are smooth, the horizon generators enter the horizon at points of caustic.  However, since the caustic locus forms a set of measure zero on the horizon, in the following discussion we will neglect this subtlety.} that are 
entirely contained within the event horizon.  
Let us choose coordinates $(\lambda,\alpha^a)$, with $a=1, \cdots ,d-1$, on the event horizon  such that $\alpha^a$ are constant along these null geodesics and $\lambda$ is a future  directed parameter (not necessarily affine) along the geodesics. As 
$\partial_{\lambda}$ is 
orthogonal to every other tangent vector on the manifold including itself, 
it follows that the metric restricted on the event horizon takes the form

\begin{equation} \label{hormetf}
ds^2= \, g_{ab}\,  d\alpha^a \, d\alpha^b
\end{equation}
Let $g$ represent the determinant of the $(d-1)\times (d-1)$ metric  $g_{ab}$.
We define the entropy $(d-1)$-form as the appropriately normalized area form on 
the spatial sections of the horizon\footnote{This definition is consistent with the Noether charge derivation of entropy currents, {\it a la} Wald, \cf., \cite{Iyer:1994ys} for a discussion for dynamical horizons. We review the connection with Wald's construction briefly in \App{wald-app}.}
\begin{equation} \label{enttf1} 
a=\frac{1}{4\, G_{d+1}} \, \sqrt{g} \, d\alpha^1\wedge d\alpha^2 \wedge \ldots \wedge d\alpha^{d-1}
\end{equation}
The area increase theorems of general relativity\footnote{We assume here that the null energy condition is satisfied. This is true of the Lagrangian used in  \cite{Bhattacharyya:2007jc} to construct the gravitation background \req{formmet}.} 
are tantamount to the monotonicity of the function $g$, \ie, 
\begin{equation}\label{ginc}
\frac{\partial g}{\partial \lambda} \geq 0
\end{equation}
which of course leads to 
\begin{equation} \label{cltf}
da= \frac{\partial_\lambda \sqrt{g}}{4\, G_{d+1}} \, 
d\lambda \wedge d\alpha^1 \wedge d\alpha^2 \ldots \wedge d\alpha^{d-1} \geq 0 \ . 
\end{equation}
We have chosen here an orientation on the horizon $\CH$ by declaring a $d$-form to be positive if it is a positive multiple of the $d$-form
$d\lambda \wedge d\alpha^1 \wedge d\alpha^2 \ldots \wedge d\alpha^{d-1} $. 

\subsection{Entropy $(d-1)$-form in global coordinates}
\label{sec:globcoord}
The entropy $(d-1)$-form described above was presented in a
special set of $\alpha^a$ coordinates which are well adapted to the horizon. 
We will now evaluate this expression in terms of a more general 
set of coordinates. Consider a set of  coordinates $x^\mu$ for the spacetime in 
the neighbourhood of the event horizon, chosen so that surfaces of 
constant $x^0=v$ intersect the horizon on spacelike slices $\Sigma_v$. 
The  coordinates $x^\mu$ used in \eqref{formmet} provide an example of such a 
coordinate chart (as we will see these are valid over a much larger range than the neighbourhood of the horizon).

As surfaces of constant $v$ are spacelike, the null geodesics that generate 
the event horizon each intersect any of these surfaces exactly once. 
Consequently, we may choose the coordinate $v$ as 
a parameter along geodesics. Then we can label the geodesics by $\alpha^a$, the 
value of $x^a$ at which the geodesic in question intersects the surface 
$v=0$. The coordinate system $\{v, \alpha^a\}$ is of the form described 
in \sec{sec:areaf}; as a result  in these coordinates  the entropy $(d-1)$-form is given by \eqref{enttf1}.  We will now rewrite this  expression in terms of the coordinates $x^\mu$ at $v=0$; for this purpose  we need the formulas for the change of coordinates from $x^\mu$ to  $\{v,\alpha^a\}$, in a neighbourhood of $v=0$. It is easy to verify that 
\begin{equation}\label{coordchange} 
\begin{split}
x^a& = \alpha^a+ \frac{n^a}{n^v}\, v + \frac{v^2}{2 \,n^v}\, 
 n^\mu\,\partial_\mu \left( \frac{n^a}{n^v}\right) + {\cal O}(v^3) \cdots \\
d x^a&=d \alpha^a + v \,d \alpha^k \, \partial_k \left( 
\frac{n^a}{n^v} \right) +  d v \,\left( \frac{n^a}{n^v} + \frac{v}{n^v}  \,
n^\mu\,\partial_\mu\left( \frac{n^a}{n^v} \right)\right) + {\cal O}(v^2) 
  \, 
\end{split}
\end{equation}

 The coordinate transformation \eqref{coordchange} allows us to write an expression for the metric on the event horizon in terms  of the coordinates  $\{v,\alpha^a\}$, in a neighbourhood of $v=0$. Let $H_{\mu \nu} \,dx^\mu \, dx^\nu  = G_{MN}\, dx^M\, dx^N|_\CH$ denote the metric restricted to the event horizon  in the $x^\mu$ coordinates.
\begin{equation}\label{metexp}
\begin{split}
ds_\CH^2 &= H_{\mu\nu}(x)\, dx^\mu \,dx^\nu \equiv g_{ab} \,d\alpha^a 
\,d\alpha^b\\
&=
h_{i j} \! \left(v,\alpha^i+\frac{n^i}{n^v}\right) \, 
\left( d \alpha^i +v \, d \alpha^k \, \partial_k  \left( \frac{n^i}{n^v}
\right) \right)
\left( d \alpha^j +v\, d \alpha^k \,\partial_k  \left( \frac{n^j
}{n^v}
\right) \right) +{\cal O}(v^2) 
\end{split}
\end{equation}
where $h_{ij}(v,x)$ is the restriction of the metric $H_{\mu\nu}$ onto a spatial slice $\Sigma_v$, which is a constant-$v$ slice.
Note that since the horizon is null,
all terms with explicit factors of $dv$ 
cancel from \eqref{metexp} in line with the general expectations presented 
in \sec{sec:areaf}.  It follows that the determinant of the induced metric, $\sqrt{g}$ of \eqref{enttf1}, is given  as
\begin{equation}\label{detmet}
\sqrt{g}= \sqrt{h} + \frac{v}{n^v}  \, 
\left( n^i \,\partial_i \sqrt{h} + \sqrt{h}\, n^v \, \partial_i \frac{n^i}{n^v} 
\right) +{\cal O}(v^2) \ , 
\end{equation}
where $h$ is the determinant of the metric on $\Sigma_v$, in $x^\mu$ 
coordinates (restricted to $v=0$). 

We are now in a position to  evaluate the area $(d-1)$-form   
\begin{equation}\label{enttf2}
a=\frac{\sqrt{h}}{4\,G_{d+1}} \, d\alpha^1\wedge d\alpha^2 \ldots \wedge d \alpha^{d-1} \ , 
\end{equation}
at $v=0$. Clearly, for this purpose we can simply set to zero all terms in \eqref{coordchange} with explicit powers of $v$, which implies that $d\alpha^a= dx^a-\frac{n^a}{n^v}\, dv$ and
\begin{equation}\label{enttf3}
a=\frac{\sqrt{h}}{4\,G_{d+1}} \,  \left(  dx^1\wedge dx^2 \ldots \wedge d x^{d-1}
-\sum_{i=1}^{d-1}\, \frac{n^i}{n^v} \, d \lambda \wedge dx^1 \wedge \ldots \wedge dx^{i-1} \wedge dx^{i+1} \wedge \ldots  \wedge dx^{d-1} \right) 
\end{equation}
From \eqref{enttf3} we can infer that the area-form can be written in terms a current as
\begin{equation}\label{enttf4} 
a=\frac{\epsilon_{\mu_1 \mu_2 ... \mu_{d}}}{(d-1)!} \; 
J_S^{\mu_1} \, dx^{\mu_2} \wedge ...\wedge dx^{\mu_{d}}
\end{equation}
where $J_S^\mu$ is given by  
\begin{equation}\label{encur}
J^\mu_S= \frac{\sqrt{h}}{4\,G^{(d+1)}_N}\,  \frac{n^\mu}{n^v}
\end{equation}
 and  our choice of orientation leads to $\epsilon_{v12\cdots (d-1)}=1$. 
In \App{wald-app}, we show that one can obtain this expression using the construction of an entropy $(d-1)$-form due to Wald, see \eqref{wald-dyn-area}.
 We can further establish that 
\begin{equation}\label{enttfd} 
d a=\frac1{(d-1)!}\epsilon_{\mu_1 \mu_2 \, \ldots \,  \mu_{d}}\, 
\partial_\alpha J_S^{\alpha} \, dx^{\mu_1} \wedge ...\wedge dx^{\mu_{d}}
\end{equation}
so that $da$ is simply the flat space Hodge dual of $\partial_\mu J_S^\mu$. 
While the appearance of the flat space Hodge dual might be puzzling at first 
sight, given the non-flat metric on $\CH$,  its origins will become clear 
once we recast this discussion in terms of the fluids dynamical variables.

\subsection{Properties of the area-form and its dual current}
\label{sec:lorinv}
Having derived the expression for the area-form we pause to record some properties which will play a role in interpreting $J^\mu_S$ as an entropy current in hydrodynamics.
 
\paragraph{Non-negative divergence:} 
Firstly, we note that the positivity of $da$ (argued for on general grounds 
in \sec{sec:areaf}) guarantees the positivity  of $\partial_\mu J^\mu_S$; 
hence we have $\partial_\mu \, J^\mu_S  \ge 0$. This in fact may be verified 
algebraically from \req{detmet}, as
\begin{equation}\label{detmetn}
\frac1{4G_{d+1}}\partial_v (\sqrt{g}) = \partial_\mu J^\mu_S \ . 
\end{equation}
The positivity of $\partial_v (\sqrt{g})$ thus guarantees that of $\partial_\mu J^\mu_S$ as is expected on general grounds.

\paragraph{Lorentz invariance:} 
The final result for our entropy current, \eqref{enttfd},  is invariant under Lorentz transformations of the coordinate $x^\mu$ (a  physical requirement of the entropy current for relativistic fluids) even though this is not  manifest. We now show that this is indeed the case.
 
Let us boost to coordinates  $\hat{x}^\mu= \Lambda^{\ \mu}_\nu x^\nu$; denoting the horizon metric in the new  coordinates by $\hat{h}_{\mu\nu}$ and the boosted normal vector by $\hat{n}^\mu$ we find 
\begin{equation} \label{transfmet} 
h_{ij}=A_i^{\ m} \, A_j^{\ n} \,\hat{h}_{mn}, \qquad A_i^{\ m}= \Lambda_i^{\ m}-\frac{\Lambda_i^{\ v} \, \hat{n}^m}
{\hat{n}^v}
\end{equation}
(where we have used $\hat{n}^\mu \hat{h}_{\mu\nu}=0$ ). It is not difficult to verify that 
$$\det A= \frac{ \left( \Lambda^{-1} \right)^{\ v}_\mu n^\mu }{\hat{n}^v}
=\frac{n^v}{\hat{n}^v}$$ 
from which it follows that $\frac{\sqrt{h}}{n^v}=\frac{\sqrt{\hat{h}}}{\hat{n}^v}$, thereby proving that our area-form defined on the a spatial section of the horizon is indeed Lorentz invariant.

\section{The Horizon to Boundary Map}
\label{sec:bdymap}

\subsection{Classification of ingoing null geodesics near the boundary}
\label{sec:nbdygds}

Our discussion thus far has been an analysis of the causal structure
of the spacetime described by the metric in   \eqref{formmet} and the
construction of an area-form on spatial sections of the horizon in
generic spacetimes. As we are interested in transporting information
about the entropy from the horizon to the boundary (where the fluid
lives), we need to define a map between the boundary and the horizon. The
obvious choice is to map the point on the boundary with coordinates
$x^\mu$ to the point on the horizon with coordinates $(r_H(x),x^\mu)$. 
More geometrically, this corresponds to moving along the geodesics
$x^\mu={\rm constant}$.
However, congruences of null geodesics shot
inwards from the boundary of AdS are far from unique. Hence, we
digress briefly to present a  characterization of the most general
such congruence. In \sec{sec:tchoice} we will then see how the
congruence of geodesics with constant $x^\mu$ fits into this general
classification. 

We will find it simplest to use Fefferman-Graham 
coordinates to illustrate our point.   Recall that any asymptotically AdS$_{d+1}$ spacetime may be put in the form  
\begin{equation}\label{gf}
ds^2=\frac{du^2+ \left( \eta_{\mu\nu}+ 
u^d \,{\phi_{\mu\nu}(w)} \right) dw^\mu\, dw^\nu} {u^2} \ , 
\end{equation}
in the neighbourhood of the boundary.   The collection of null geodesics that 
intersect the boundary point  $(w^\mu, u=0)$ are given by the equations 
\begin{equation}\label{nullgeo}
\frac{dw^A}{d\lambda}=u^2\, \left(t^A + {\cal O}\left(u^d\right)\right)
\end{equation}
where $A$ runs over the $d+1$ variables $\{u, w^\mu\}$ and the null 
tangent vector  must obey $t^A\, t_A=0$. It is always possible to re-scale the affine parameter 
to set $t^u=1$; making this choice, our geodesics are labelled by a $d$-vector $t^\mu$ satisfying $\eta_{\mu \nu}\, t^\mu\,  t^\nu=-1$. With these conventions $t^\mu$ may be regarded as a $d$-velocity. In summary, the set of ingoing null geodesics that emanate from any given boundary point are parameterized by the $d-1$ directions in which they can go -- this parameterization is conveniently 
encapsulated in terms of a unit normalized timelike $d$-vector $t^\mu$ which 
may, of course, be chosen as an arbitrary function of $x^\mu$. Consequently, 
congruences of ingoing null geodesics are parameterized by an arbitrary
 $d$-velocity field, $t^\mu(x)$ on the boundary of AdS.

\subsection{Our choice of $t^\mu(x)$}
\label{sec:tchoice}

It is now natural to ask  what $t^\mu(x)$ is for the congruence
defined by $x^\mu=const$ in the coordinates of \cite{Bhattacharyya:2007jc}. 
The answer to this question is easy to work out, and turns out to be 
satisfyingly simple: for this choice of congruence, $t^\mu(x)=u^\mu(x)$ 
where $u^\mu(x)$ is the velocity field of fluid dynamics!\footnote{ In 
order to see this note that 
\begin{equation}\label{translation} \begin{split}
u_\mu \frac{dx^\mu}{d \lambda}& =  u_\mu \,\frac{dw^\mu}{d \lambda} 
+\frac{du}{d \lambda} \\
P_{\nu \mu} \frac{dx^\mu}{d \lambda}& = {\cal P}_{\nu \mu} 
\frac{dw^\mu}{d \lambda} \\
\end{split}
\end{equation} 
whereas indicated quantities on the LHS of \eqref{translation} refer to the coordinate system of \cite{Bhattacharyya:2007jc}, the quantities on the RHS refer to the Fefferman-Graham  coordinates \eqref{gf}. It follows from these formulae that the geodesic with  $t^A = (1, u^\mu)$ maps to the null geodesic $\frac{dx^\mu}{d\lambda}=0$ in the coordinates used to write \eqref{formmet}.}

While metrics dual to fluid dynamics are automatically equipped with a 
velocity field, it is in fact also possible to associate a velocity field 
with a much larger class of asymptotically AdS spacetimes. Recall that any such 
spacetime has a boundary stress tensor $T_{\mu\nu}$.\footnote{In a general 
coordinate system the stress tensor is proportional to the extrinsic curvature
of the boundary slice minus local counter-term subtractions. In the Fefferman-Graham coordinate system described above, the final answer is especially 
simple; $T_{\mu\nu} \propto \phi_{\mu \nu}(x^\mu)$.} For most such spacetimes 
there is a natural velocity field associated with this stress tensor; the 
velocity $u^\mu(x)$ to which one has to boost in order that $T^{0i}$ vanish 
at the point $x$. More invariantly, $u^\mu(x)$ is chosen to be the unique 
timelike eigenvector of the matrix $T^\mu_{\ \nu}(x)$.\footnote{This prescription
breaks down when $u^\mu$ goes null - \ie,\ if there exist points at which 
the energy moves at the speed of light.}
That is, we choose $u^\mu(x)$ to satisfy 
\begin{equation}\label{velsolve}
(\eta_{\mu \nu}+u_\mu\, u_\nu) \,  T^{\nu \kappa}\, u_\kappa=0 
\end{equation}
This definition of $u^\mu(x)$ coincides precisely with the 
velocity field in \cite{Bhattacharyya:2007jc} (this is the so-called 
Landau frame). The null congruence given by 
$t^\mu(x)=u^\mu(x)$ is now well defined for an arbitrary asymptotically 
AdS spacetime, and reduces to the congruence described earlier in this section 
for the metrics dual to fluid dynamics.

\subsection{Local nature of the event horizon}
\label{sec:horloc}

As we have seen in \sec{sec:horizon} above, the event horizon is effectively 
local for the metrics dual to fluid dynamics such as \eqref{formmet}. In particular, the position  of the event horizon $r_H(x^\mu)$ depends only on the values and derivatives of the fluid  dynamical variables in a neighbourhood of $x^\mu$ and not elsewhere in 
spacetime. Given the generic teleological behaviour of event horizons (which requires knowledge of the entire future evolution of the spacetime), this  feature 
of our event horizons is rather unusual. To shed light on this issue, we supply an intuitive explanation for this phenomenon, postponing the  actual evaluation of the function $r_H(x^\mu)$ to \sec{sec:locfm}.  

The main idea behind our intuitive explanation may be stated rather 
simply. As we have explained above, the metric of \cite{Bhattacharyya:2007jc}
is tube-wise well approximated by tubes of the metric of a uniform black 
brane at constant velocity and temperature. Now consider a uniform black brane 
whose parameters are chosen as $u_\mu=(-1,0,0,0)$ and $b=1/(\pi \,T)=1$ by 
a choice of coordinates. In this metric a radial outgoing null 
geodesic that starts at  $r=1 + \delta$ (with $\delta \gg \epsilon$) and 
$v=0$ hits the boundary at a time  $\delta v=\int \frac{dr}{r^2 f(r)} 
\approx -4 \,\ln \delta$. Provided this radial outgoing geodesic
well approximates the path of a geodesic in the metric of 
\cite{Bhattacharyya:2007jc} throughout its trajectory, it follows that the 
starting point of this geodesic lies outside the event horizon of the 
spacetime. 

The two conditions for the approximation described above to be valid are: 
\begin{enumerate} 
\item That geodesic in question lies within the tube in which the metric
of \cite{Bhattacharyya:2007jc} is well approximated by a black brane with 
constant parameters throughout its trajectory. This is valid when $\delta v
\approx  -4 \,\ln \delta \ll 1/\epsilon$. 
\item That even within this tube, the small corrections to the metric of 
\cite{Bhattacharyya:2007jc} do not lead to large deviations in the geodesic. 
Recall that the radial geodesic in the metric of \cite{Bhattacharyya:2007jc}
is given by the equation $$\frac{dv}{dr}=-\frac{G_{rv} + \CO\left(\epsilon\right) }{G_{vv}+\CO\left(\epsilon\right)}
=\frac{2 +\CO\left(\epsilon\right) }{f(r)+ \CO\left(\epsilon\right)} .$$
This geodesic well approximates that of the uniform black brane provided 
the $\CO\left(\epsilon\right)$ corrections above are negligible, a 
condition that is met  provided $f(r) \gg \epsilon$, 
\ie, when $|r-1|= \delta \gg \epsilon$.
\end{enumerate}

Restoring units we conclude that a point at $r=\frac{1}{b}(1+ \delta)$ 
necessarily lies outside the event horizon provided $\delta \gg \epsilon$ 
(this automatically ensures  $\delta v \approx  -4 \,\ln \delta \ll 1/\epsilon$.
when $\epsilon$ is small). 

In a similar fashion it is easy to convince oneself that all geodesics that are emitted from $r=\frac1{b}(1-\delta)$ hit  the singularity within the regime of validity of the tube approximation provided  $\delta \gg \epsilon$. Such a point therefore lies inside the event horizon.  It follows that the event horizon in the solutions of \cite{Bhattacharyya:2007jc} is given by  the hypersurface $r= \pi\, T \,( 1 + \CO\left(\epsilon\right))$.

\section{Specializing to Dual Fluid Dynamics}
\label{sec:fluid}

We will now proceed to determine the precise form of the event 
horizon manifold to second order in $\epsilon$ using the results obtained in \sec{sec:horizon}.
This will be useful to construct the entropy current in the fluid dynamics utilizing the map derived in \sec{sec:bdymap}.

\subsection{The local event horizon dual to fluid dynamics}
\label{sec:locfm}

The metric dual to fluid flows given in \cite{Bhattacharyya:2007jc} takes the form 
\eqref{formmet} with explicitly determined  
forms of the functions in that metric (see \App{app:notation}). We list the properties and values of these functions that we will need below:\footnote{Since we require only the values of the functions appearing in the metric \eqref{csdefn} and \eqref{chidef} at $ r= 1/b$ to evaluate \eqref{secondorder}, we present here the functions evaluated at this specific point. The full expressions can be found in \App{app:notation}, see \eqref{sabdefs} and \eqref{jdefns}.}
\begin{equation}\label{listing} \begin{split}
f(1) &= 0  , \qquad s_a^{(1)}  = 0 , \qquad s_c^{(1)}  = 0, \\
s_b^{(1)}&=\frac{2}{3}\, \frac{1}{b} \,\partial_\mu u^\mu  \ , \qquad  \partial_r s_b^{(1)}=\frac{2}{3}\, \partial_\mu u^\mu \ ,\\
j_\mu^{(1)}&=-\frac{1}{b} \,u^\nu\, \partial_\nu u_\mu \ , \qquad
t^{(1)}_{\mu\nu}=\frac{1}{b}\left(\frac{3}{2}\ln 2 + \frac{\pi}{4}\right)  \,  \sigma_{\mu\nu} \equiv F\,\sigma_{\mu\nu} \\
s_a^{(2)}& =\frac{3}{2}\, s_c^{(2)}= \frac{b^2}{16}\left(2 \, \ST{4} - \ST{5}\, \left( 2 + 12\, {\cal C} +  \pi + \pi^2 - 9 \, (\ln 2)^2 - 3\pi\, \ln 2 + 4\, \ln 2\right)\right)\\
s_b^{(2)}& = -\frac{2}{3} \,\Ss{3}  + \ST{1} -\frac{1}{9}\,\ST{3} - \frac{1}{12}\,\ST{4} + \ST{5}\,\left(\frac{1}{6} +  {\cal C} + \frac{\pi}{6} + \frac{5 \,\pi^2}{48} + \frac{2}{3}\ln 2 \right) \\
j_\mu^{(2)}& = \frac{1}{16}{\bf B}^{\infty} - \frac{1}{144}{\bf B}^{\rm fin} \\
\end{split}
\end{equation}
where ${\cal C}$ is the Catalan number. We encounter here various functions (of the boundary coordinates) which are essentially built out the fluid velocity $u^\mu$ and its derivatives. These have been abbreviated to symbols such as  $\Ss{3}$, $\ST{1}$, \etc, and  are defined \eqref{scalrep}. Likewise ${\bf B}^{\infty}$ and ${\bf B}^{\rm fin}$ are defined in \eqref{bidefs}.

Using the equation for the conservation of stress tensor ($\partial_\mu T^{\mu\nu} = 0$) up to second order in derivatives one can simplify the expression for $r_H$ \eqref{secondorder}.
Conservation of stress tensor gives
\begin{equation}\label{constress}
\partial_\nu\left[\frac{1}{b^4}\, \left(\eta^{\mu\nu} + 4 u^\mu u^\nu\right)\right] = 
\partial_\nu\left[\frac{2}{b^3}\sigma^{\mu\nu}\right]
\end{equation}
Projection of \eqref{constress} into the co-moving and transverse directions, achieved by contracting it with $u_\mu$ and $P_{\mu\nu}$ respectively, we find
\begin{equation}\label{simp}
\begin{split}
s_b^{(1)} - \frac{2}{b^2} \, u^\mu \, \partial_\mu b &= \frac{1}{3}\, \sigma_{\mu\nu}\, \sigma^{\mu\nu} = \CO\left(\epsilon^2\right)\\
P^{\mu\nu} \left(b^2\, j_{\mu}^{(1)} + \partial_\mu b\right) &= -\frac{b^2}{2}\, P^{\ \nu}_\mu \left(\partial_{\alpha}\sigma^{\alpha\mu} - 3\,\sigma^{\mu\alpha} \, u^\beta\, \partial_\beta u_\alpha\right) + \CO\left(\epsilon^3\right)
\end{split}
\end{equation}
Inserting \eqref{simp} into \eqref{firstorder} we see that $r_{(1)}$ of 
\eqref{firstorder} simply 
vanishes for the spacetime dual to fluid dynamics, and so, to first order in $\epsilon$,  
$r_H^{(1)}= \frac{1}{b}$. At next order this formula is corrected to  
\begin{equation}\label{simanto}
r_H^{(2)}  = \frac{1}{b(x)} + r_{(2)}(x)= \frac{1}{b} + \frac{b}{4}\left(s_b^{(2)}+\frac{1}{3} \, \sigma_{\mu\nu} \, \sigma^{\mu\nu}\right)
\end{equation}
In order to get this result we have substituted into \eqref{secondorder} 
the first of \eqref{simp}, utilized the fact that $r_{(1)} =0$ and the observation 
(from the second line of \eqref{simp}) that 
$$P^{\mu\nu}\,\left(b^2 j_{\mu}^{(1)} + \partial_\mu b\right)\, 
\left(b^2 j_{\nu}^{(1)} + \partial_\nu b\right)= {\cal O}(\epsilon^4) $$

In this special case the components of  normal vector in the boundary directions \eqref{nm1} (accurate to $\CO\left(\epsilon^2\right)$) are given by
\begin{equation}\label{nm2} 
n^\mu = \left( 1+ s_a^{(2)} \right)u^\mu  -\frac{b^2}{2}\, P^{\mu\nu}\, \left(\partial^{\alpha}\sigma_{\alpha\nu} - 3\,\sigma_{\nu\alpha}\, u^\beta\, \partial_\beta u^\alpha\right) 
+ b^2 \, P^{\mu \nu} \, j_\nu^{(2)}  \ .
\end{equation}
%

\subsection{Entropy current for fluid dynamics}
\label{sec:finecur}

We will now specialize the discussion of \sec{sec:globcoord} to the metric of 
\cite{Bhattacharyya:2007jc}, using the formulae derived in \sec{sec:locfm}.  
In the special case of the metric of \cite{Bhattacharyya:2007jc} we have 
\begin{equation}\label{sg}\begin{split}
\sqrt{g} &= \frac{1}{b^3} \left( 1-\frac{b^4}{4}\, F^2 \, \sigma_{\mu \nu}\, \sigma^{\mu\nu} 
+3 \,b \, r_{(2)} + s_a^{(2)} \right)\\
=& \frac{1}{b^3} \left( 1-\frac{b^4}{4}F^2 \, \sigma_{\mu \nu}\, \sigma^{\mu\nu} +\frac{b^2}{4} \sigma_{\mu\nu}\,  \sigma^{\mu\nu}
+ \frac{3 \,b^2}{4} \,s_b^{(2)} + s_a^{(2)} \right) \ ,
\end{split}
\end{equation}
where the various quantities are defined in \eqref{listing}. We conclude from \eqref{encur} that
\begin{equation}\label{ansj} \begin{split}
4 \, G^{(5)}_N\, b^3 \, J_S^\mu  &=  u^\mu \,  \left( 1-\frac{b^4}{4} \, F^2 \, \sigma_{\alpha \beta}\, \sigma^{\alpha \beta}+\frac{b^2}{4} \sigma_{\alpha \beta} \, \sigma^{\alpha \beta} +\frac{3 \, b^2}{4} \, s_b^{(2)} + s_a^{(2)} \right)\\
& + b^2 \, P^{\mu \nu} \left[ -\frac{1}{2}\left( \partial^{\alpha}\sigma_{\alpha\nu} - 3 \,\sigma_{\nu\alpha} \,u^\beta \,\partial_\beta u^\alpha\right) +  j_\nu^{(2)} \right] \ .
\end{split}
\end{equation}
This is the expression for the fluid dynamical entropy current which we derive from the gravitational dual.
\section{Divergence of the Entropy Current}
\label{sec:positive}

In previous sections, we have presented a gravitational construction
of an entropy current which, we have argued, is guaranteed to have
non-negative divergence at each point. We have also presented an explicit
construction of the entropy current to order $\epsilon^2$ in the
derivative expansion. In this section we directly compute the
divergence of our entropy current and verify its positivity. We will
find it useful to first start with an abstract analysis of the most
general Weyl invariant entropy current in fluid dynamics and compute
its divergence, before specializing to the entropy current constructed
above.

\subsection{The most general Weyl covariant entropy current and its divergence}
\label{sec:genweyl}

The entropy current in $d$-dimensions has to be a Weyl covariant vector of weight $d$. We will work in four dimensions ($d=4$) in this section, and so will consider currents that are Weyl covariant vector of weight 4. Using the equations 
of motion, it may be shown that there exists a 7 dimensional family of 
two derivative weight 4 Weyl covariant vectors that have the 
correct equilibrium limit for an entropy current. In the 
notation of \cite{Loganayagam:2008is}, (reviewed in \App{app:weylcov}),  
this family may be parameterized as 
\begin{equation}\label{derivexpjsmain:eq}
\begin{split}
(4\pi\,\eta)^{-1}\,J^\mu_S = 4 \,G^{(5)}_N \,b^3 \,J^\mu_S = &\left[1+b^2\left(A_1 \,\sigma_{\alpha\beta}\,\sigma^{\alpha\beta}+A_2 \,\omega_{\alpha\beta}\,\omega^{\alpha\beta} +A_3 \,\mathcal{R}\,\right)\right]u^\mu \\
&\quad  + b^2 \,\left[ B_1 \,\mathcal{D}_\lambda \sigma^{\mu\lambda} + B_2 \,\mathcal{D}_\lambda \omega^{\mu\lambda} \right]\\
&\quad + C_1\ b\ \l^\mu + C_2\ b^2 u^\lambda \,\mathcal{D}_\lambda \l^\mu  +\ldots 
\end{split}
\end{equation}
where $b=(\pi T)^{-1},\eta=(16\pi \, G^{(5)}_N \, b^3)^{-1}$ and the rest of the notation is as in \cite{Loganayagam:2008is} (see also \App{app:notation} and \App{app:weylcov}). 

In Appendix \ref{app:weylcov} we have computed the divergence of this entropy current 
(using the third order equations of motion derived and expressed in Weyl covariant language in \cite{Loganayagam:2008is}). Our 
final result is 
\begin{equation}\label{divjsmain:eq}
\begin{split}
4 \,G^{(5)}_N \,b^3 \,\mathcal{D}_\mu J^\mu_S &= \frac{b}{2}\,\left[\sigma_{\mu\nu}+ b \,\left(2 \,A_1+4 \,A_3-\frac{1}{2}+\frac{1}{4}\,\ln 2\right)\, u^\lambda\mathcal{D}_\lambda \sigma^{\mu\nu}+4\,b \,(A_2+A_3)\,\omega^{\mu\alpha} \omega_{\alpha}{}^{\nu}\right.\\
&\qquad\left. + b \,(4 \,A_3-\frac{1}{2})(\sigma^{\mu\alpha} \,\sigma_{\alpha}{}^{\nu} )+b\, C_2 \, \mathcal{D}^\mu \l^\nu \right]^2\\
&\quad + (B_1-2A_3)\, b^2\, \mathcal{D}_\mu \mathcal{D}_\lambda \sigma^{\mu\lambda} +(C_1+C_2) \,b^2\, \l_\mu \mathcal{D}_\lambda \sigma^{\mu\lambda} +\ldots\\
\end{split}
\end{equation}

Note that the leading order contribution to the divergence of the arbitrary 
entropy current is proportional to $\sigma_{\mu\nu} \, \sigma^{\mu\nu}$. This term is of second 
order in the derivative expansion, and is manifestly non-negative. In addition 
the divergence has several terms at third order in the derivative expansion. 

Within the derivative expansion the second order piece dominates all third 
order terms whenever it is nonzero. However it is perfectly possible for 
$\sigma_{\mu\nu}$  to vanish at a point -- $\sigma_{\mu\nu}$ are simply 5 of several independent  Taylor coefficients in the  expansion  of the velocity field at a point (see \App{app:inddata} for details). When that happens the third order terms 
are the leading contributions to $\CD_\mu J^\mu_S$. Since such
terms are cubic in derivatives they are odd orientation reversal ($x^\mu \rightarrow 
-x^\mu$), and so can be non-negative for all velocity configurations 
only if they vanish identically. We conclude that positivity requires 
that the RHS of \eqref{divjsmain:eq} vanish upon setting $\sigma_{\mu\nu}$
to zero.

As is apparent, all terms on the first two lines of \eqref{divjsmain:eq} are 
explicitly proportional to $\sigma_{\mu\nu}$. The two independent expressions on the third line 
of that equation are in general nonzero even when $\sigma_{\mu\nu}$ 
vanishes. As a result $\mathcal{D}_\mu J^\mu_S \ge 0$ requires that 
the second line of \eqref{divjsmain:eq} vanish identically; hence, we obtain the 
following constraints on coefficients of the second order terms in the 
entropy current 
\begin{equation}\label{condnmain:eq}
\begin{split}
B_1=2 \, A_3 \qquad&\qquad C_1+C_2=0 \\
\end{split}
\end{equation}
for a non-negative divergence entropy current. 

These two conditions single out a 5 dimensional submanifold of non-negative 
divergence entropy currents in the 7 dimensional space 
\eqref{derivexpjsmain:eq} of candidate Weyl covariant entropy currents. 

Since a local notion of  entropy is an emergent thermodynamical construction 
(rather than a first principles microscopic construct), it seems reasonable 
that there exist some ambiguity in the definition of a local entropy 
current. We do not know, however, whether this physical ambiguity is 
large enough to account for the full 5 parameter non uniqueness described 
above, or whether a physical principle singles out a smaller sub family of 
this five dimensional space as special. Below we will see that our 
gravitational current - which is special in some respects - may be generalized
to a two dimensional sub family in the space of positive divergence currents. 

\subsection{Positivity of divergence of the gravitational entropy current}
\label{sec:posgec}

It may be checked (see \App{app:weylcov}) that our entropy current \eqref{ansj} 
may be rewritten 
in the form \eqref{derivexpjsmain:eq} with the coefficients 
\begin{equation}\label{ABCmain:eq}
\begin{split}
A_1 = \frac{1}{4}+\frac{\pi}{16}+\frac{\text{ln}\ 2}{4}  ;\qquad A_2 &=-\frac{1}{8} ;\qquad A_3 =\frac{1}{8} \\
B_1=\frac{1}{4} ;\qquad& B_2=\frac{1}{2}  \\
C_1 = &\; C_2 = 0 \\
\end{split}
\end{equation} 
 It is 
apparent that the coefficients listed in \eqref{ABCmain:eq} obey the constraints of 
positivity \eqref{condnmain:eq}. This gives a direct algebraic check 
of the positivity 
of the divergence of $\eqref{ansj}$.

The fact that it is possible to write the current \eqref{ansj} in the form 
\eqref{derivexpjsmain:eq} also demonstrates the Weyl covariance of 
our current \eqref{ansj}.

\subsection{A two parameter class of gravitational entropy currents}
\label{sec:gravamb}

As we have seen above,  there exists a five parameter set of non-negative 
divergence conformally covariant entropy currents that have the correct 
equilibrium  limit. An example of such a current was first 
constructed in \cite{Loganayagam:2008is}.

Now let us turn to an analysis of possible generalizations 
of the gravitational entropy current presented in this paper. Our 
construction admits two qualitatively distinct,  reasonable sounding, 
generalizations that we now discuss. 

Recall that we constructed our entropy $(d-1)$-form via the pullback of 
the area-form on the event horizon. While the area-form is a very natural 
object, all its physically important properties (most importantly the 
positivity of divergence) appear to be retained if we add to it the 
exterior derivative of a $(d-2)$-form. This corresponds to the addition of
the exterior derivative of a $(d-2)$-form to the entropy current $J^\mu_S$. Imposing the 
additional requirement of Weyl invariance at the two derivative level this 
appears to give us the freedom to add a multiple of $\frac{1}{b} \,{\mathcal D}_\lambda  
\omega^{\lambda \sigma}$  to the entropy current in four dimensions.

In addition, we have the freedom to modify our boundary to horizon map in 
certain ways; our construction of the entropy current \eqref{ansj} depends on this map and we have made the specific choice described in \sec{sec:bdymap}. Apart from geometrical 
naturalness and other aesthetic features, our choice had two important 
properties. First,  under this map $r_H(x^\mu)$ (and hence the local 
entropy current) was a local function of the fluid dynamical variables at 
$x^\mu$. Second, our map was Weyl covariant; in particular, the entropy 
current obtained via this map was automatically Weyl covariant. We will now 
parameterize all boundary to horizon maps (at appropriate order in the
derivative expansion) that preserve these two desirable properties. 

Any one to one boundary to horizon map may be thought of as a boundary to 
boundary diffeomorphism compounded with the map presented in \sec{sec:bdymap}. 
In order to preserve the locality of the entropy current, this 
diffeomorphism must be small (\ie, of sub-leading order 
in the derivative expansion). At the order of interest, 
it turns out to be sufficient to study diffeomorphisms 
parameterized by a vector $\delta\zeta$ that is of at most first order in the 
derivative expansion. In order that our entropy current have 
acceptable Weyl transformation properties under this map, $\delta\zeta$ must 
be Weyl invariant. Up to terms that vanish by the equations of motion, this 
singles out a two parameter set of acceptable choices for $\delta\zeta$; 
\begin{equation}\label{zetapar}
\delta\zeta^\mu = 2\ \delta\lambda_1\ b\ u^\mu + \delta\lambda_2\ b^2\ \l^\mu
\end{equation}

To leading order the difference between the $(d-1)$-forms obtained by pulling the area $(d-1)$-form $a$ back under the two different maps is  given by the Lie derivative of the pull-back $s$ of $a$  
$$\delta s= {\cal L}_{\delta\zeta} \,s= d(\delta\zeta _\mu \,s^\mu) + \delta\zeta_\mu\, (d s)^\mu. $$ 
Taking the boundary Hodge dual of this difference we find 
\begin{equation}\label{entambig}
\begin{split}
\delta J^\mu_S  &= \mathcal{L}_{\delta\zeta} J^\mu_S -  J^\nu_S\, \nabla_\nu \delta\zeta^\mu \\
&= \mathcal{D}_\nu\left[J^\mu_S \,\delta\zeta^\nu- J^\nu_S\, \delta\zeta^\mu \right] + \delta\zeta^\mu \,\mathcal{D}_\nu J^\nu_S
\end{split}
\end{equation}
 Similarly 
\begin{equation} \label{dissip} \begin{split}
\delta \partial_\mu J_s^\mu &= \delta\zeta^\mu\,\partial_\mu \partial_\nu J_s^\nu + \partial_\mu\delta\zeta^\mu\;
\partial_\nu J_s^\nu = {\cal L}_{\delta\zeta} \,\partial_\mu J_s^\mu + \partial_\mu \delta\zeta^\mu \,   \partial_\nu J_s^\nu \\
&={\cal L}_{\delta\zeta} \,\mathcal{D}_\mu J_s^\mu + \mathcal{D}_\mu 
\delta\zeta^\mu \,   \mathcal{D}_\nu J_s^\nu
\end{split}
\end{equation}
Using the fluid equations of motion it turns that the RHS of $\eqref{entambig}$ 
is of order $\epsilon^3$ (and so zero to the order retained in this paper) 
for $\zeta^\mu \propto b^2\, l^\mu$. Consequently, to second order we find a 
one parameter generalization of the entropy current -- resulting from the 
diffeomorphisms \eqref{zetapar} with $\delta \lambda_2$ set to zero. 

Note that, apart from the diffeomorphism shift,  the local rate of entropy 
production changes in magnitude (but not in sign) under redefinition 
\eqref{entambig} by a 
factor proportional to the Jacobian of the coordinate transformation 
parameterized by $\delta\zeta$. In Appendix \ref{app:weylcov} we  have explicitly computed the 
shift in the current \eqref{ansj} under the operation described in 
\eqref{entambig} (with $\delta\zeta$ of the form \eqref{zetapar}) and also 
explicitly verified the invariance of the positivity of divergence under 
this map. 

In summary we have constructed a two parameter generalization of our 
gravitational entropy current \eqref{ansj}. One of these two parameters 
arose from the freedom to add an exact form to the area form on the horizon. 
The second parameter had its origin in the freedom to generalize the 
boundary to horizon map.

\section{Discussion}
\label{sec:discuss}

We have demonstrated that any singularities in the metrics of 
\cite{Bhattacharyya:2007jc}, dual to fluid dynamics, are shielded behind 
a regular event horizon (we expect the same to be true for the solution of 
\cite{VanRaamsdonk:2008fp}). Further, we have shown that the structure of 
this event horizon is determined locally by the variables of fluid 
dynamics, and have presented an explicit expression for the location of 
the event horizon to second order in the $\epsilon$ (boundary derivative) 
expansion. Remarkably, the event horizon, which is a global concept in 
general relativity, turned out to be rather simple to locate, partly due 
to our choice of particularly useful coordinate system \req{formmet}, and 
more importantly due to the long-wavelength requirement that our solution 
be dual to a system described by fluid dynamics.  We emphasize that within 
the boundary derivative expansion of this paper we are directly able to 
construct the event horizon; we did not need to discuss other more local 
constructs like the apparent horizon as an intermediate step towards 
understanding the global structure of our solutions.

We have also constructed an entropy $(d-1)$-form on the 
event horizon of an arbitrary $d+1$ dimensional spacetime and used the \
pullback of this form to the boundary to construct a manifestly non-negative 
divergence entropy current for asymptotically AdS$_{d+1}$ solutions of 
gravity with a horizon. We have derived an explicit expression for this 
entropy current for the solutions dual to \cite{Bhattacharyya:2007jc} and 
demonstrated a direct algebraic check of the positivity of divergence of 
this current within fluid dynamics. 

In order to lift the entropy $(d-1)$-form from the horizon to the boundary, we used a natural map between the horizon and the boundary, given by ingoing null geodesics which emanate from the boundary in the direction of the fluid flow.  These ingoing geodesics in fact determine the coordinate system of 
\cite{Bhattacharyya:2007jc} (they constitute lines of constant $x^{\mu}$ for the metric \eqref{formmet}). 

We also directly studied a seven parameter family of weight four Weyl covariant 
fluid dynamical vectors that have the appropriate equilibrium limit to be 
an entropy current and demonstrated that a 5 parameter subclass of 
this family of currents has non negative divergence to second order in 
the derivative expansion. The entropy currents we constructed via a pullback 
of the area form constitute a special subclass of these currents. It is 
natural to inquire what the gravitational interpretation of the remaining 
currents is.\footnote{We thank M. Van Raamsdonk for raising this question.}
It is natural to wonder whether they are assosicated with 
appararent horizons\footnote{We thank A. Strominger for stressing the 
physical relevance of apparent horizons to our situation, and for a very 
useful related discussion.} and other quasi-local horizons (such as trapping/dynamical horizons, isolated horizons, {\it e.g.},\cite{Booth:2005qc,Ashtekar:2004cn}). At least several of these 
horizons also appear to obey versions of the area increase theorem. 
Consequently, it should be possible to obtain conserved entropy currents 
via the pullback of a suitably defined area form on these horizons. 
Apparent and other dynamical horizons have one initially unpalatable feature;
their structure depends on a choice of  the slicing of spacetime into 
spacelike surfaces. However perhaps it is precisely this ambiguity that 
allows these constructions to cover the full 5 parameter set of non negative
entropy currents discussed above?\footnote{A cautionary note is in order;
it is possible that a subclass of the 5 parameter non negative divergence 
entropy currents is an artifact of the derivative expansion, and has no
continuation to finite $\epsilon$. We thank M. Van Raamsdonk for discussions
on this point.} Note that in the context of dynamical horizons, \cite{Gourgoulhon:2006uc} obtain\footnote{We thank I. Booth for bringing this reference to our attention.}  a characterization of a membrane fluid obeying non-relativistic hydrodynamics equations with a uniquely specified entropy. Their system has rather different characteristics (absence of shear-viscositly for instance) and appears to model the black hole as a fluid, rather than construct an explicit dual as in the current discussion. It would interesting to understand this connection better.

We re-emphasize that our results demonstrate that each of the solutions 
of \cite{Bhattacharyya:2007jc} (with regular fluid data) has its singularities
hidden from the boundary by a regular event horizon. Consequently all 
gravitational solutions dual to regular solutions of 
fluid dynamics obey the cosmic censorship conjecture. It would be interesting
to investigate how our results generalize to irregular (\eg, turbulent) 
solutions of fluid dynamics, as also to gravitational solutions beyond 
the long wavelength expansion. As we have explained in the Introduction, 
several such solutions are dual descriptions of the field 
theoretic approach towards local equilibrium. The appearance of a naked 
singularity in this approach would appear to imply singularities of 
real time correlation functions in this process. It would be fascinating 
to study this connection in more detail. On a more speculative, or perhaps 
more ambitious note, it is natural to inquire what (if any) feature of 
field theoretic correlators would be sensitive to the apparently crazy 
nature of near singularity dynamics even when the latter is cloaked by 
a horizon. 

Another interesting direction concerns $\alpha'$  corrections to the bulk solutions, which in the language of the dual field theory correspond to finite 't Hooft coupling effects. In the bulk, there is a well developed formalism due to Wald \cite{Wald:1993nt,Iyer:1994ys} which provides 
a generalization of the Bekenstein-Hawking area formula for the entropy of the black hole to higher derivative gravity. The main idea is to construct the entropy of (asymptotically flat) solutions using a variational principle of the Lagrangian; essentially, from the variational principle one obtains the first law of thermodynamics, which is used  to construct the entropy as a Noether charge. This construction of the Noether charge entropy is conceptually similar to the area-form we present and in fact reduces to it in the two derivative limit.  It might be possible
-- and would be very interesting -- to generalize the discussion presented 
in this paper to be able to account for $\alpha'$ corrections (see
Appendix \ref{wald-app}). The key  issue here would be to find an analogue of the  area increase theorem for $\alpha'$ corrected gravity. This is complicated from a pure general relativity standpoint, owing to the fact that higher derivative theories generically violate the energy conditions.\footnote{In the context of supersymmetric solutions in $\alpha'$ corrected gravity, as discussed in \cite{Hubeny:2004ji} for the so called small black holes, while the entropy remains proportional to the area and specific classes of solutions satisfy the averaged null energy condition \cite{Borde:1987qr}; one still is unable to show the desired monotonicity property of entropy.} The AdS/CFT  correspondence seems to require that such a generalization exist, and it  would be very interesting to determine it. It is possible that the  requirement of the existence of such a theorem provides `thermodynamical' 
constraints to $\alpha'$ corrections of the low energy equations of 
gravity.\footnote{In \cite{Jacobson:1994qe} the authors demonstrate the second law for the Einstein Hilbert action deformed by an $R^2$ term; this Lagrangian however is not likely to arise as the low energy effective action from string theory \cite{Zwiebach:1985uq}. See also  \cite{Brigante:2008gz,Brigante:2007nu} for recent discussions of constraints on parameters appearing in higher derivative theories in connection to the dual hydrodynamic description.}

If the gravity/fluid dynamics correspondence could be understood in more 
detail for confining gauge theories (see \eg, \cite{Aharony:2005bm,Lahiri:2007ae}),
fluid dynamics could give us a handle on very interesting horizon dynamics. 
For instance, one might hope to explore the possibility of topological transitions of the event horizon.\footnote{While Cosmic Censorship precludes splitting 
of black holes, they can easily merge without the curvatures becoming large.}

Turning to more straightforward issues, it would be interesting to generalize 
the discussion of this paper to encompass the study of field theory arbitrary 
curved manifolds. It would also be interesting to generalize our analysis to the bulk dual of charged fluid flows, and especially to the flows of 
extremal charged fluids.  This could permit more direct contact with 
the entropy functional formalism for extremal black holes.  A natural framework for  such analysis, specifically in relation to the horizon dynamics studied here, is provided by the near-horizon  
metrics for degenerate horizons discussed in \cite{Kunduri:2007vf}.
Furthermore, one might hope that such a study could have bearing on studying the
behaviour of superfluids.

Finally, note that while field theoretic conserved currents are most naturally 
evaluated at the boundary of AdS, the entropy current most naturally lives 
on the horizon. This is probably related to the fact that while field 
theoretic conserved currents are microscopically defined, the notion of 
a local entropy is an emergent long distance concept, and so naturally 
lives in the deep IR region of geometry, which, by the UV/IR map, 
is precisely the event horizon.
Correspondingly, we find it fascinating that, in the limits studied in this paper, 
the shape of the event horizon is a local reflection of fluid variables. 
This result is reminiscent of the membrane paradigm of black hole physics. 
It would be fascinating to flesh out this observation, and perhaps to 
generalize it.

\subsection*{Acknowledgements}

We would like to acknowledge useful discussions and correspondences with 
J. Bhattacharya, I. Booth, B. Freivogel, R. Gopakumar, S. Hartnoll, M. Headrick, G. Horowitz, P. Joshi, J. Lucietti, A. Strominger, S. Trivedi, M. Van Raamsdonk,  S. Wadia,  T. Wiseman and
all students in the TIFR theory room. The work 
of SM was supported in part by a Swarnajayanti Fellowship. VH and MR
are supported in part by STFC. HSR is a Royal Society University
Research Fellow. SB, RL, GM, and SM  and would also like to
acknowledge our debt to the people of  India for their generous and
steady support to research in the basic sciences.

\appendix
\section{Notation}
\label{app:notation}

We work in the mostly positive, $(-++\ldots)$, signature.  The dimensions of the spacetime in which the conformal fluid lives is denoted by $d$. In the context of AdS/CFT, the dual AdS$_{d+1}$ space has $d+1$ spacetime dimensions.  The event horizon is a $d$-dimensional null manifold $\mathcal{H}$. $\mathcal{H}$ is foliated by $d-1$ dimensional constant $v$ spatial slices denoted by $\Sigma_v$. The induced metric on $\Sigma_v$ is denoted by $h_{ab}$ (and $h$ denotes its determinant).

Latin alphabets $A,B,\ldots$ are used to denote the $d+1$ dimensional bulk indices which range over $\{r,0,1,\ldots,d-1\}$. Lower Greek letters $\mu,\nu,\ldots$ indices range over $\{0,1,\ldots,d-1\}$ and lower case Latin letters $a,b,\ldots$ indices range over $\{1,\ldots,d-1\}$.   The co-ordinates in the bulk are denoted by $X^A$ which is often split into a radial co-ordinate $r$ and $x^\mu$. We will often split $x^\mu$ into $v$ and $x^a$. 

In these co-ordinates, the equation for the horizon takes the form  $\mathcal{S}_{\mathcal{H}}\equiv r-r_{H}(x) =0$. We can choose to eliminate the co-ordinate $r$ in favour of $x^\mu$'s via this equation. Then, in the $x^\mu$ co-ordinates the components of the metric are denoted by $H_{\mu\nu}$. In addition, we find it convenient to use a co-ordinate system  ${\alpha^a,\lambda}$  on $\mathcal{H}$ -- in these co-ordinates, the components of the induced metric on the horizon take the special form $g_{\lambda\lambda}= g_{a \lambda} =0$ and $g_{ab}\neq 0$. 

Our convention for the Riemann curvature tensor is fixed by the relation
\begin{equation}
[\nabla_\mu,\nabla_\nu]V^\lambda=R_{\mu\nu\sigma}{}^{\lambda}V^\sigma . 
\end{equation}

In Table \ref{nottable}, we list the physical meaning and the definitions of various quantities used in the text, referring to the equations defining them where appropriate:

\begin{table}[h]
  \begin{center}
  \begin{tabular}{||c|l||c|l||}
    \hline
    {\bf Symbol} & {\bf Definition} & {\bf Symbol} & {\bf Definition} \\
    \hline
    $d$ & dimensions of boundary & $\mathcal{H}$& The event horizon (d-dimensional)\\
    $X^A$  & Bulk co-ordinates & $x^\mu$ & Boundary co-ordinates\\
    $\Sigma_v$ & A spatial slice of $\mathcal{H}$ & $\lambda,\alpha^a$ & Co-ordinates on $\mathcal{H}$\\
    $G_{AB}$ & Bulk metric,\eqref{formmet} & $\eta_{\mu\nu}$ & Boundary metric (Minkowski) \\
    $h_{ab}$ & Induced metric on $\Sigma_v$ &$g_{ab}$ & Metric on $\Sigma_v \subset\mathcal{H}$  \\
    $r_H(x)$ & Horizon function, \eqref{ehsurf}& $H_{\mu\nu}$ & Induced metric on $\CH$\\
    $\mathcal{S}_{\mathcal{H}}=0$ & Eqn. of Horizon & $s$& Entropy (d-1)-form on $\Sigma_\lambda$ \\
    $\xi^A$ & Normal vector to the Horizon \eqref{nadef}& $ n^\mu$ &  See \eqref{nxidef}\\    
\hline \hline
    $s_a^{(k)}$& See \eqref{formmet}, \eqref{sabdefs}, \eqref{listing} &$s_b^{(k)}$& See \eqref{formmet}, \eqref{sabdefs}, \eqref{listing} \\
    $j_\mu^{(k)}$&See \eqref{formmet}, \eqref{jdefns}, \eqref{listing} &$t_{\mu\nu}^{(k)}$ &See \eqref{formmet}, \eqref{listing}\\
    \hline \hline
%
    $T$ & Fluid temperature & $\eta$ & Shear viscosity  \\
    $T^{\mu\nu}$ & Energy-momentum tensor & $J^\mu_S$ & Entropy current \\
    $u^\mu$ & Fluid velocity ($u^\mu u_\mu =-1$) &  $P^{\mu\nu}$ & Projection tensor, $
    \eta^{\mu\nu}+u^\mu u^\nu$ \\
    $a^\mu$ & Fluid acceleration, \eqref{fderdefs}  &$\vartheta$&  Fluid expansion,  \eqref{fderdefs}\\
    $\sigma_{\mu\nu}$ & Shear strain rate, \eqref{fderdefs} & $\omega_{\mu\nu}$ & Fluid vorticity, \eqref{fderdefs} \\
    $\pi_{\mu\nu}$ & Visco-elastic stress  & & \\
    \hline \hline
%
     $\mathcal{D}_\mu$ & Weyl-covariant derivative  & $\mathcal{A}_\mu$ & See  \eqref{weylcov:eq}\\
     $R_{\mu\nu\lambda}{}^{\sigma}$ & Riemann tensor &      $\mathcal{F}_{\mu\nu}$ & $\nabla_\mu\mathcal{A}_\nu-\nabla_\nu\mathcal{A}_\mu$ \\
     $R_{\mu\nu},R$ & Ricci tensor/scalar & $\mathcal{R_{\mu\nu}},\mathcal{R}$& See \eqref{weylcov:eq} \\
     $G_{\mu\nu}$ & Einstein tensor & $\mathcal{G}_{\mu\nu}$ & See \eqref{weylcov:eq}\\
     $C_{\mu\nu\lambda\sigma}$ & Weyl curvature & &\\
     \hline
\end{tabular}
\end{center}
\caption{Conventions used in the text}
\label{nottable}
\end{table}
\subsection{Fluid dynamical parameters}

Various expressions in the text and are built out of the fluid velocity; we list them here for convenience. The basic building blocks are the derivatives of the fluid velocity, decomposed into appropriate representations based on their symmetries. We have (see the table above for the physical meaning of these parameters),   
\begin{equation}
\begin{split}
\vartheta &= \partial_{\mu}u^{\mu} \\
 a^{\nu}&= u^{\mu}\, \partial_{\mu}u^{\nu} \\
 \sigma^{\mu\nu} &= \frac{1}{2} \left( P^{\lambda \mu} \partial_\lambda  u^{\nu}+  P^{\lambda \nu} \partial_\lambda u^{\mu} \right)  - \frac{1}{3} \, P^{\mu\nu}\,\partial_{\lambda}u^{\lambda} \\
\omega^{\mu \nu} &= \frac{1}{2} P^{\mu \alpha}\, P^{\nu \beta}\, 
\left( \partial_{\alpha}u_{\beta} - \partial_{\beta}u_{\alpha} \right) \\ 
\l^\mu &= \epsilon^{\alpha\beta\nu\mu}\omega_{\alpha\beta} u_\nu \\
\end{split}
 \label{fderdefs}
\end{equation}

In addition, we will have occasion at various points in the text to encounter various functions built out of the first derivatives of the fluid velocity defined in \eqref{fderdefs}. These functions were defined in \cite{Bhattacharyya:2007jc} to present the second order metric, and show up for example in \eqref{listing}. 
We have:
\begin{equation} \label{scalrep}
\begin{split}
& \Ss{3} = \frac{1}{b} 
P^{\alpha \beta}\,\partial_\alpha \partial_\beta b\, \qquad 
\ST{1}  = \CD u^\alpha \, \CD u_\alpha \ , \qquad 
\ST{2} =  \l_\mu \, \CD u^\mu \\
&\ST{3} = (\partial_\mu u^\mu)^2\ , \qquad 
\ST{4} = \l_\mu \,\l^\mu\ , \qquad
\ST{5} = \sigma_{\mu\nu}\,\sigma^{\mu\nu} \ .
\end{split}
\end{equation}
where $\CD = u^\mu \, \partial_\mu$ (Note that this is a different derivative from the Weyl covariant derivative introduced in \App{app:weylcov}; the distinction should be clear from the context).
\begin{equation} \label{vecrep}
\begin{split}
& \V{4}_\nu=\frac{9}{5}\left[ \frac{1}{2}P^\alpha_\nu \, P^{\beta\gamma}\,\partial_\gamma\, \left( \partial_{\beta} u_{\alpha}  +\partial_{\alpha} u_{\beta} \right) -  \frac{1}{3} \,P^{\alpha\beta}\, P^\gamma_{\;\nu}\, \partial_\gamma  \partial_\alpha u_\beta \right] -  P^\mu_ \nu\, P^{\alpha\beta}\, \partial_\alpha\partial_\beta u_{\mu}\\
& \V{5}_\nu = P^\mu_ \nu\, P^{\alpha\beta}\partial_\alpha\partial_\beta u_{\mu}\\
&\VT{1}_\nu =\partial_\alpha u^\alpha \, \CD u_\nu \ , \qquad 
\VT{2}_\nu =\epsilon_{\alpha\beta\gamma\nu} \, u^\alpha\, \CD u^\beta \,\l^\gamma\ , \qquad 
\VT{3}_\nu= \sigma_{\alpha\nu}\, \CD u^\alpha\ .
\end{split}
\end{equation}
%

\subsection{The functions appearing in the second order metric}

The metric \eqref{formmet} derived in \cite{Bhattacharyya:2007jc}  has been rewritten in terms of various auxiliary functions used to define $\CS(r, x^\mu)$ and
$\chi_{\mu \nu}(r, x^\mu)$. These functions can be read off from Eq (5.25) of  \cite{Bhattacharyya:2007jc}; we list them here for convenience.\footnote{ One notational change we have made is to rename the functions $\alpha^{(k)}_{\mu \nu}$ appearing in \cite{Bhattacharyya:2007jc} to $t^{(k)}_{\mu \nu}$. We don't list this here as it doesn't appear directly in our analysis of the entropy current. }

\paragraph{Scalars under $SO(3)$ spatial rotations:} The scalar functions appearing at first and second order are respectively,
\begin{equation}\label{sabdefs}
\begin{split}
s_a^{(1)}(r,x^\mu) &= 0  \\
s_a^{(2)}(r,x^\mu) &= \frac{3}{2}\, b^2 \, h_2(b\, r) \\
s_b^{(1)}(r,x^\mu) & = \frac{2}{3}\,  r\, \partial_{\lambda} u^{\lambda}\\
s_b^{(2)}(r,x^\mu) & = \frac{1}{r^2} \,\frac{k_2(b\, r)}{b^2}\\
\end{split}
\end{equation}
in terms of several functions of $r$ which are given as
\begin{equation}
\label{fdef}
F(r)  =\frac{1}{4}\, \left[\ln\left(\frac{(1+r)^2(1+r^2)}{r^4}\right) - 2\,\arctan(r) +\pi\right] 
\end{equation}	
Defining
\begin{equation}\label{}
\begin{split}
S_h(r) & \equiv\frac{1}{3\, r^3}\, \ST{4} + \frac{1}{2}\, W_h(r) \, \ST{5}  \\
S_k(r)& \equiv 12 r^3\, h_2(r) + (3 \, r^4 -1)\, h_2'(r) 
 -\frac{4 \,r}{3}\, \Ss{3} + 2 \,r \,\ST{1}  -\frac{2\, r}{9}\, \ST{3}+ \frac{1 + 2\, r^4}{6\, r^3} \,\ST{4} + \frac{1}{2}\,W_k(r) \,\ST{5}\ ,
\end{split}
\end{equation}	
where the functions $W_h(r)$ and $W_k(r)$ are given by
\begin{equation*}
\begin{split}
 W_h(r) &=\frac{4}{3}\; \frac{\left(r^2+r+1\right)^2-2 \left(3 \,r^2+2 r+1\right) \, F(r)   }{ r \,\left(r+1\right)^2\,\left(r^2+1\right)^2 } \ ,\\
 W_k(r) &=\frac{2}{3}\;\frac{4\, \left(r^2+r+1\right)\,\left(3\,r^4-1\right)\, F(r) - \left(2 r^5+2 r^4+2 r^3-r-1\right)}{ r\,\left(r+1\right)\,\left(r^2+1\right) } \ .
\end{split}
\end{equation*}
 The other symbols $\Ss{3}$,  $\ST{1}$, \etc, are defined in \eqref{scalrep}. We can now write the expressions for the functions appearing in the definition of $s_{a,b}^{(2)}$ as 
\begin{equation}\label{h2solution}
\begin{split}
h_2(r) &= - \frac{1}{4\, r^2}\, S^{\infty}_h+\int_r^\infty \frac{dx}{x^5}\; \int_x^\infty dy\ y^4 \, \left(S_h(y) - \frac{1}{y^3}\, S_h^{\infty} \right) \\
k_2(r) &= \frac{r^2}{2}\, S^{\infty}_k-\int_r^\infty dx\, \left(S_k(x) - x\, S_k^{\infty}\right) \ .
\end{split}
\end{equation}
where we have defined 
\begin{equation}
 S_h^{\infty} =\left(\frac{1}{3}\, \ST{4} + \frac{2}{3}\,\ST{5} \right) \ , \qquad 
S^{\infty}_k \equiv \left(-\frac{4 }{3} \,\Ss{3} + 2  \,\ST{1}-\frac{2}{9}\, \ST{3} -\frac{1}{6}\, \ST{4}+\frac{7}{3} \,\ST{5}\right)  \ .
\end{equation}	

\paragraph{Vectors under $SO(3)$ spatial rotations:} The vector functions appearing at first and second order are respectively,
\begin{equation}
\begin{split}
j_\mu^{(1)}(r,x^\mu) & = - r\, u^{\alpha}\, P^\beta_\mu \,\partial_\alpha u_{\beta}  \\
j_\mu^{(2)}(r,x^\mu)  &= -\frac{1}{\, b^2 \, r^2} \, P^\alpha_\mu \, \left( -\frac{r^2}{36} \, {\bf B}^\infty_\alpha + \int_r^\infty dx \; x^3\; \int_x^\infty dy \,\left({\bf B}_\alpha(y) - \frac{1}{9\, y^3} \,{\bf B}^\infty_\alpha\right) \right) 
\end{split}
\label{jdefns}
\end{equation}	
where
\begin{equation} \label{bi}
{\bf B}(r) = \frac{\left(2 \,r^3+2 r^2 +2\,r -3\right)\,  {\bf B}^{\infty} + {\bf B}^{{\rm fin}}}{18\, r^3\,  (r+1) \, \left(r^2+1\right)} 
\end{equation}
with
\begin{equation}\label{bidefs}
\begin{split}
{\bf B}^\infty  &= 4\, \left(10\, \V{4} +  \V{5} +3\, \VT{1} -3 \,\VT{2} -6\,\VT{3} \right)\\
{\bf B}^{{\rm fin}} & = 9\, \left(20\,  \V{4}- 5 \,\VT{2} - 6\,\VT{3}\right) ,
\end{split}
\end{equation}	
The symbols $\V{k}$ and $\VT{k}$ are defined above in \eqref{vecrep} as derivatives of the fluid velocity.

\section{Weyl covariant formalism}\label{app:weylcov}

In this appendix, we present the various results related to Weyl covariance in hydrodynamics that are relevant to this paper. The conformal nature of the boundary fluid dynamics strongly constrains the form of the stress tensor and the entropy current \cite{Baier:2007ix,Loganayagam:2008is}. An efficient way of exploiting this symmetry is to employ a manifestly Weyl-covariant formalism for hydrodynamics that was introduced in the reference \cite{Loganayagam:2008is}. 

In brief, for an arbitrary tensor with weight $w$, one defines a Weyl-covariant derivative\footnote{In contrast to the analysis in the main text, we find it convenient here to work with an arbitrary background metric, whose associated torsion-free connection is used to define the covariant derivative $\nabla_\mu$.}
\begin{equation}\label{D:eq}
\begin{split}
\mathcal{D}_\lambda\ Q^{\mu\ldots}_{\nu\ldots} &\equiv \nabla_\lambda\ Q^{\mu\ldots}_{\nu\ldots} + w\  \mathcal{A}_{\lambda} Q^{\mu\ldots}_{\nu\ldots} \\ 
&+\left[{g}_{\lambda\alpha}\mathcal{A}^{\mu} - \delta^{\mu}_{\lambda}\mathcal{A}_\alpha  - \delta^{\mu}_{\alpha}\mathcal{A}_{\lambda}\right] Q^{\alpha\ldots}_{\nu\ldots} + \ldots\\
&-\left[{g}_{\lambda\nu}\mathcal{A}^{\alpha} - \delta^{\alpha}_{\lambda}\mathcal{A}_\nu  - \delta^{\alpha}_{\nu}\mathcal{A}_{\lambda}\right]  Q^{\mu\ldots}_{\alpha\ldots} - \ldots
\end{split}
\end{equation}
where the Weyl-connection $\mathcal{A}_\mu$ is related to the fluid velocity via the relation
\begin{equation}\label{defA:eq}
\begin{split}
\mathcal{A}_\mu = {u}^\lambda\nabla_\lambda {u}_\mu - \frac{\nabla_\lambda {u}^\lambda}{d-1} {u}_\mu
\end{split}
\end{equation}
We shall exploit the manifest Weyl covariance of this formalism to establish certain results concerning the entropy current that are relevant to the discussion in the main text.

In \sec{app:genEntCurr}, we write down the most general Weyl-covariant entropy current and compute its divergence. This computation leads us directly to an analysis of the constraints on the entropy current imposed by the second law of thermodynamics. This analysis generalizes and completes the analysis in \cite{Loganayagam:2008is} where a particular example of an entropy current which satisfies the second law was presented. Following that, in \sec{app:gravweylcov}, we rewrite the results of this paper in a Weyl-covariant form and show that the expression for the entropy current derived in this paper satisfies the constraint derived in \sec{app:genEntCurr}. This is followed by a discussion in \sec{app:ambecur} on the ambiguities in the definition of the entropy current.

\subsection{Constraints on the entropy current: Weyl covariance and the second law}\label{app:genEntCurr}

We begin by writing down the most general derivative expansion of the entropy current in terms  Weyl-covariant vectors of weight $4$.\footnote{We will restrict attention to fluid dynamics in $3+1$ dimensions.} After taking into account the equations of motion and various other identities, the most general entropy current consistent with Weyl covariance can be written as:
\begin{equation}\label{derivjs:eq}
\begin{split}
(4\pi\,\eta)^{-1}\,J^\mu_S = 4 \,G^{(5)}_N \,b^3 \,J^\mu_S = &\left[1+b^2\left(A_1 \,\sigma_{\alpha\beta}\,\sigma^{\alpha\beta}+A_2 \,\omega_{\alpha\beta}\,\omega^{\alpha\beta} +A_3 \,\mathcal{R}\,\right)\right]u^\mu \\
&\quad  + b^2 \,\left[ B_1 \,\mathcal{D}_\lambda \sigma^{\mu\lambda} + B_2 \,\mathcal{D}_\lambda \omega^{\mu\lambda} \right]\\
&\quad + C_1\ b\ \l^\mu + C_2\ b^2 u^\lambda \,\mathcal{D}_\lambda \l^\mu  +\ldots 
\end{split}
\end{equation}
where $b=(\pi \,T)^{-1}$ and we have already assumed the leading order result for the entropy density $s=4\,\pi\,\eta=(4 \,G^{(5)}_N \,b^3)^{-1}$ and $\l^\mu = \epsilon^{\alpha\beta\nu\mu}\omega_{\alpha\beta} u_\nu $.\footnote{We shall follow the notations of \cite{Loganayagam:2008is} in the rest of this appendix. In particular, we recall the following definitions 
\begin{equation}\label{weylcov:eq}
\begin{split}
\mathcal{A}_\mu = a_\mu - \frac{\vartheta}{3}u_\mu \ ;\qquad & \mathcal{F}_{\mu\nu} = \nabla_\mu \mathcal{A}_\nu - \nabla_\nu \mathcal{A}_\mu \\
\mathcal{R} = R -6 \nabla_\lambda \mathcal{A}^\lambda + 6 \mathcal{A}_\lambda \mathcal{A}^\lambda \ ; \qquad& \mathcal{D}_\mu u_\nu = \sigma_{\mu\nu} + \omega_{\mu\nu} \\
\mathcal{D}_\lambda \sigma^{\mu\lambda} = \nabla_\lambda \sigma^{\mu\lambda}- 3 \mathcal{A}_\lambda \sigma^{\mu\lambda} \ ;\qquad& \mathcal{D}_\lambda \omega^{\mu\lambda} = \nabla_\lambda \omega^{\mu\lambda}- \mathcal{A}_\lambda \omega^{\mu\lambda}  \\
\end{split}
\end{equation} 
Note that in a flat spacetime, $R$ is zero but $\mathcal{R}$ is not. Though we will always be working in flat spacetime, we will keep the $R$-terms around to make our expressions manifestly Weyl-covariant. }

Now, we want to derive the constraints imposed by the second law on the A,B and C coefficients appearing above. To this end, we take the divergence of the entropy current above to get
\begin{equation}\label{divjs1:eq}
\begin{split}
4 \,G^{(5)}_N \,b^3\, \mathcal{D}_\mu J^\mu_S = & -3 b^{-1}\,u^\mu\,\mathcal{D}_\mu b -2\, C_1\ \l^\mu\mathcal{D}_\mu b\\
&+ b^2\, \mathcal{D}_\mu\left[\left(A_1 \,\sigma_{\alpha\beta}\,\sigma^{\alpha\beta}+A_2 \,\omega_{\alpha\beta}\,\omega^{\alpha\beta} +A_3\, \mathcal{R}\right)\,u^\mu \right.\\
&\qquad\left.  +\left( B_1 \,\mathcal{D}_\lambda \sigma^{\mu\lambda} + B_2\, \mathcal{D}_\lambda \omega^{\mu\lambda} + C_2\ u^\lambda\, \mathcal{D}_\lambda \l^\mu \right) \right] +\ldots
\end{split}
\end{equation}
where we have used the facts that $\mathcal{D}_\mu \l^\mu=0$ and that $\mathcal{D}_\mu b$ gets non-zero contributions only at second order  \eqref{binvDb:eq}.  Further,  $u^\lambda \,\mathcal{F}_{\mu\lambda}$ gets non-zero contributions only at third order (the equations of motion force  $u^\lambda \mathcal{F}_{\mu\lambda}=0$ at second order). 

In order to simplify the expression further, we need the equations of motion. Let us write the stress tensor in the form
\begin{equation}\label{pidef:eq}
\begin{split}
T^{\mu\nu}= (16\pi \,G^{(5)}_N \,b^4)^{-1}\; (\eta^{\mu\nu}+4\, u^\mu u^\nu) + \pi^{\mu\nu}
\end{split}
\end{equation}
where $\pi_{\mu\nu}$ is transverse -- $u^\nu\pi_{\mu\nu}=0$. This would imply
\begin{equation}\label{binvDb:eq}
\begin{split}
0 &=b^4\,u_\mu\, \mathcal{D}_\nu  T^{\mu\nu} = 
b^4\,\mathcal{D}_\nu ( u_\mu T^{\mu\nu} ) - b^4\,(\mathcal{D}_\nu u_\mu) \,T^{\mu\nu}\\
&\Longrightarrow \;  4\,\left(\frac{3}{b}\, u^\mu\mathcal{D}_\mu b - \frac{b}{4\,\eta} \,\sigma_{\mu\nu} \pi^{\mu\nu}\right) = 0\\
\end{split}
\end{equation}
where we have multiplied the equation by $ 16\pi \,G^{(5)}_N $ in the second line to express things compactly.
Similarly, we can write $2\,\l^\mu\,\mathcal{D}_\mu b= -b^2 \,\l_\mu \mathcal{D}_\lambda \sigma^{\mu\lambda}$ which is exact upto third order in the derivative expansion. Note that these 
are just the Weyl-covariant forms of the equations that we have already encountered in \eqref{simp}.

We further invoke the following identities(which follow from the identities proved in the Appendix A of \cite{Loganayagam:2008is}) \footnote{Since we are only interested in the case where boundary is conformally flat, we will consistently neglect terms proportional to the Weyl curvature in the following.}
\begin{equation}\label{divscalar:eq}
\begin{split}
\mathcal{D}_\mu (\sigma_{\alpha\beta}\,\sigma^{\alpha\beta} \,u^\mu) &= 2 \,\sigma_{\mu\nu}\,u^\lambda\mathcal{D}_\lambda \sigma^{\mu\nu}\\
\mathcal{D}_\mu (\omega_{\alpha\beta}\,\omega^{\alpha\beta} \,u^\mu) &= 4 \sigma^{\mu\nu}\,\omega_\mu{}^\alpha \,\omega_{\alpha\nu} -2 \,\mathcal{D}_\mu \mathcal{D}_\lambda \omega^{\mu\lambda}\\
\mathcal{D}_\mu (\mathcal{R}\, u^\mu) &= -2\,\sigma_{\mu\nu}\,\mathcal{R}^{\mu\nu}+ \mathcal{D}_\mu\left[-2 \,\mathcal{D}_\lambda \sigma^{\mu\lambda} +2\, \mathcal{D}_\lambda \omega^{\mu\lambda}+ 4 \,u_\lambda \,\mathcal{F}^{\mu\lambda}\right] \\
-2\,\sigma_{\mu\nu}\mathcal{R}^{\mu\nu}&= 4 \,\sigma_{\mu\nu}\,\left[u^\lambda\mathcal{D}_\lambda \sigma^{\mu\nu}+\omega^{\mu\alpha} \,\omega_{\alpha}{}^{\nu}+\sigma^{\mu\alpha}\, \sigma_{\alpha}{}^{\nu}-C^{\mu\alpha\nu\beta}\,u_\alpha u_\beta \right]\\
\mathcal{D}_\mu(u^\lambda \,\mathcal{D}_\lambda \l^\mu) &= \mathcal{D}_\mu(\l^\lambda\,\mathcal{D}_\lambda u^\mu)-\mathcal{F}_{\mu\nu}\,\l^\mu u^\nu \\
\mathcal{D}_\mu(\l^\lambda \,\mathcal{D}_\lambda u^\mu) &= \sigma_{\mu\nu}\,\mathcal{D}^\mu \l^\nu + \l_\mu \,\mathcal{D}_\lambda \sigma^{\mu\lambda}\\
\end{split}
\end{equation}
to finally obtain
\begin{eqnarray}\label{divjs2:eq}
4 \,G^{(5)}_N \,b^3 \,\mathcal{D}_\mu J^\mu_S \!\!&=&  b^2\, \sigma_{\mu\nu}\,\left[- \frac{\pi^{\mu\nu}}{4\,\eta\, b} + 2\, A_1\, u^\lambda\,\mathcal{D}_\lambda \sigma^{\mu\nu}+4 \,A_2\,\omega^{\mu\alpha}\, \omega_{\alpha}{}^{\nu}
-2 \,A_ 3\, \mathcal{R}^{\mu\nu}+C_2  \,\mathcal{D}^\mu \l^\nu \right]
\nonumber\\
&&\qquad + (B_1-2A_3)\, b^2 \,\mathcal{D}_\mu \mathcal{D}_\lambda \sigma^{\mu\lambda} +(C_1+C_2) \,b^2\,\l_\mu \,\mathcal{D}_\lambda \sigma^{\mu\lambda} +\ldots \nonumber\\
&&\hspace{-1.6cm}= b^2\sigma_{\mu\nu}\left[- \frac{\pi^{\mu\nu}}{4\,\eta\, b} + (2 A_1+4 A_3) \,u^\lambda\mathcal{D}_\lambda \sigma^{\mu\nu}
+4\, (A_2+A_3)\,\omega^{\mu\alpha}\, \omega_{\alpha}{}^{\nu}+4 \,A_3\,\sigma^{\mu\alpha} \,\sigma_{\alpha}{}^{\nu} +C_2  \,\mathcal{D}^\mu \l^\nu \right] \nonumber\\
\qquad&& + (B_1-2A_3)\, b^2\, \mathcal{D}_\mu \mathcal{D}_\lambda \sigma^{\mu\lambda} +(C_1+C_2) \,b^2 \,\l_\mu \mathcal{D}_\lambda \sigma^{\mu\lambda} +\ldots
\end{eqnarray}
Substituting the value of $\pi^{\mu\nu}$ as calculated from the known stress tensor, we find
\begin{eqnarray}\label{divjs2:eqB}
4 \,G^{(5)}_N \,b^3 \,\mathcal{D}_\mu J^\mu_S &=& b^2\sigma_{\mu\nu}\left[\frac{\sigma^{\mu\nu}}{2\, b} + \left(2\, A_1+4\, A_3-\frac{1}{2}+\frac{1}{4}\,\ln 2\right)\, u^\lambda\mathcal{D}_\lambda \sigma^{\mu\nu}\right.
\nonumber\\
&&\qquad\left.
 +\,4\, (A_2+A_3)\,\omega^{\mu\alpha} \,\omega_{\alpha}{}^{\nu} + (4 \,A_3-\frac{1}{2})\,(\sigma^{\mu\alpha} \,\sigma_{\alpha}{}^{\nu} )+C_2\, \mathcal{D}^\mu \l^\nu \right]
 \nonumber\\
&&\qquad + (B_1-2A_3) \,b^2 \,\mathcal{D}_\mu \mathcal{D}_\lambda \sigma^{\mu\lambda} +(C_1+C_2) \,b^2\, \l_\mu\, \mathcal{D}_\lambda \sigma^{\mu\lambda} +\ldots
\end{eqnarray}
This expression can in turn be rewritten in a more useful form by isolating the terms that are manifestly non-negative:
\begin{eqnarray}\label{divjspos:eq}
4 \,G^{(5)}_N \,b^3 \,\mathcal{D}_\mu J^\mu_S &=& \frac{b}{2}\,\left[\sigma_{\mu\nu}+ b \,\left(2 \,A_1+4 \,A_3-\frac{1}{2}+\frac{1}{4}\,\ln 2\right)\, u^\lambda\mathcal{D}_\lambda \sigma^{\mu\nu}+4\,b \,(A_2+A_3)\,\omega^{\mu\alpha} \omega_{\alpha}{}^{\nu}\right. \nonumber \\
&&\qquad\left. +\; b \,(4 \,A_3-\frac{1}{2})(\sigma^{\mu\alpha} \,\sigma_{\alpha}{}^{\nu} )+b\, C_2 \, \mathcal{D}^\mu \l^\nu \right]^2 \nonumber\\
&&\qquad + (B_1-2A_3)\, b^2\, \mathcal{D}_\mu \mathcal{D}_\lambda \sigma^{\mu\lambda} +(C_1+C_2) \,b^2\, \l_\mu \mathcal{D}_\lambda \sigma^{\mu\lambda} +\ldots
\end{eqnarray}

The second law requires that the right hand side of the above equation be positive semi-definite at every point in the boundary. First, we note from \eqref{divjspos:eq} that the first two lines are positive semi-definite whereas the terms in the third line are not -- given a velocity configuration in which the third line evaluates to a particular value, as argued in the main text, we can always construct another configuration to get a contribution with opposite sign. Consider, in particular, points in the boundary where $\sigma_{\mu\nu}=0$ -- at such points, the contribution of the first two lines become subdominant in the derivative expansion to the contribution from the third line. The entropy production at these points can be  positive semi-definite only if the combination the coefficients appearing in the third line vanish identically.    

Hence, we conclude that the second law gives us two constraints relating A,B and C, \viz,
\begin{equation}\label{condnB:eq}
\begin{split}
B_1=2 \, A_3 \qquad&\qquad C_1+C_2=0 \\
\end{split}
\end{equation}
Any entropy current which satisfies the above relations constitutes a satisfactory proposal for the entropy current from the viewpoint of the second law.

One simple expression for such an entropy current which satisfies the above requirements was proposed in \cite{Loganayagam:2008is}. The $J^\lambda_s$ proposed there is given by 
\begin{equation}\label{proposedjs:eq}
\begin{split}
(4\pi\,\eta)^{-1} \, J^\lambda_S &= u^\lambda -\frac{b^2}{8} \,\left[\left(\ln 2\, \sigma^{\mu\nu} \sigma_{\mu\nu}+\omega^{\mu\nu} \, \omega_{\mu\nu}\right) \, u^\lambda+2 \, u_\mu \,(\mathcal{G}^{\mu\lambda}+\mathcal{F}^{\mu\lambda})+6 \,\mathcal{D}_\nu\omega^{\lambda\nu}\right] +\ldots
\end{split}
\end{equation}
Now, using the identity
\begin{equation}\label{confident:eq}
\begin{split}
u_\mu \,(\mathcal{G}^{\mu\lambda}+\mathcal{F}^{\mu\lambda})= -\frac{\mathcal{R}}{2} \, u^\lambda-\mathcal{D}_\nu\sigma^{\lambda\nu} - \mathcal{D}_\nu\omega^{\lambda\nu}+2 \, u_\mu\mathcal{F}^{\lambda\mu}
\end{split}
\end{equation}
and the equations of motion, we can rewrite the above expression in the form appearing in \eqref{derivjs:eq} to  get the value of A,B and C coefficients as
\begin{equation}
\begin{split}\label{proposedABCeq}
A_1 = -\frac{\text{ln}\ 2}{8} ;\qquad A_2 &=-\frac{1}{8} ;\qquad A_3 = \frac{1}{8} \\
B_1=\frac{1}{4} ;\qquad& B_2=- \frac{1}{2} \\
C_1=&\; C_2  = 0 \\
\end{split}
\end{equation}
It can easily be checked that these values satisfy the constraints listed in \eqref{condnB:eq}. Further, for these values, the divergence of the entropy current simplifies considerably and we 
get
\begin{equation}\label{proposedivjs:eq}
\begin{split}
4 \, G^{(5)}_N \, b^3 \,\mathcal{D}_\mu J^\mu_S &=  \frac{b}{2} \,\sigma_{\mu\nu} \,\sigma^{\mu\nu}
\end{split}
\end{equation}

However, as the analysis in this section shows, this proposal is just one entropy current among a class of entropy currents that satisfy the second law. This is not surprising, since (as was noted in \cite{Loganayagam:2008is}) the second law alone cannot determine the entropy current uniquely. 

\subsection{Entropy current and entropy production from gravity}\label{app:gravweylcov}

We now calculate the coefficients $A_i$'s and $B_i$'s for the actual entropy current calculated from gravity in \eqref{ansj} and check whether the they obey the constraints in \eqref{condnB:eq}. Unlike the proposal in \cite{Loganayagam:2008is} , the entropy current derived in \sec{sec:fluid} takes into account the detailed microscopic dynamics(of which hydrodynamics is an effective description) encoded in the dual gravitational description.

In order to cast the entropy current in the form given by \eqref{derivjs:eq} , we have to first rewrite the quantities appearing in this paper in a Weyl-covariant form. We have the following relations in the flat spacetime which identify the Weyl-covariant forms appearing in the second-order metric of \cite{Bhattacharyya:2007jc} --
\begin{equation}\label{covforms:eq}
\begin{split}
\ST{4} = 2 \, \omega_{\alpha\beta} \, \omega^{\alpha\beta} ;\qquad& \ST{5} = \sigma_{\alpha\beta}  \,\sigma^{\alpha\beta}; \\
-\frac{4}{3}  \, \Ss{3} + 2 \, \ST{1} -\frac{2}{9} \,\ST{3} &= \frac{2}{3}  \,\sigma_{\alpha\beta}  \,\sigma^{\alpha\beta}-\frac{2}{3} \, \omega_{\alpha\beta} \omega^{\alpha\beta}+ \frac{1}{3} \,\mathcal{R} \\
\frac{5}{9}\, \V{4}_\mu + \frac{5}{9} \, \V{5}_\mu +\frac{5}{3}  \,\VT{1}_\mu-\frac{5}{12} \, \VT{2}_\mu-\frac{11}{6} \, \VT{3}_\mu &= P^{\nu}_{\mu}\, \mathcal{D}_\lambda \sigma_{\nu}{}^{\lambda}  \\
\frac{15}{9}\,\V{4}_\mu - \frac{1}{3} \,\V{5}_\mu -\VT{1}_\mu-\frac{1}{4} \, \VT{2}_\mu+\frac{1}{2}\, \VT{3}_\mu &= P^{\nu}_{\mu}\, \mathcal{D}_\lambda \omega_{\nu}{}^{\lambda} \\
\end{split}
\end{equation}
These can be used to obtain 
\begin{equation}\label{covB:eq}
\begin{split}
\mathbf{B}^{\infty}_{\mu} &= 18\, P^{\nu}_{\mu}  \,\mathcal{D}_\lambda \sigma_{\nu}^{\lambda}+18  \,P^{\nu}_{\mu} \, \mathcal{D}_\lambda \omega_{\nu}^{\lambda}\\
&= 18 \,\left(-\sigma_{\alpha\beta}  \,\sigma^{\alpha\beta}+\omega_{\alpha\beta}  \,\omega^{\alpha\beta}\right)\, u_\mu +
18  \,\mathcal{D}_\lambda \sigma_{\mu}{}^{\lambda}+18  \,\mathcal{D}_\lambda \omega_{\mu}{}^{\lambda}\\
\mathbf{B}^{\text{fin}}_{\mu}&= 54\, P^{\nu}_{\mu} \,\mathcal{D}_\lambda \sigma_{\nu}^{\lambda}+90 \,P^{\nu}_{\mu}\, \mathcal{D}_\lambda \omega_{\nu}^{\lambda}\\
&= \left(-54\,\sigma_{\alpha\beta} \,\sigma^{\alpha\beta}+90\,\omega_{\alpha\beta} \,\omega^{\alpha\beta}\right)\,u_\mu +
54  \,\mathcal{D}_\lambda \sigma_{\mu}{}^{\lambda}+90 \, \mathcal{D}_\lambda \omega_{\mu}{}^{\lambda}\\
\end{split}
\end{equation}
Hence, all the second-order scalar and the vector contributions to the metric can be written in terms of three Weyl-covariant scalars $ \sigma_{\alpha\beta} \,\sigma^{\alpha\beta}$, $\omega_{\alpha\beta} \,\omega^{\alpha\beta}$ and $\mathcal{R}$ and two Weyl-covariant vectors $ \mathcal{D}_\lambda \sigma_{\mu}{}^{\lambda}$ and $\mathcal{D}_\lambda \omega_{\mu}{}^{\lambda} $.

Using the above expressions, we can rewrite the second order scalar and the vector contributions to the entropy current appearing in  \eqref{listing} as
\begin{equation}\label{scalvecjscov:eq}
\begin{split}
s_a^{(2)} &=\frac{3}{2}\, s_c^{(2)} = -\frac{b^2}{4}\,\left(\frac{1}{2}+\text{ln}\ 2 + 3\ \mathcal{C}+\frac{\pi}{4}+\frac{5\pi^2}{16}-\left(\frac{3}{2}\text{ln}\ 2 +\frac{\pi}{4}\right)^2\right)\, \sigma_{\alpha\beta} \,\sigma^{\alpha\beta} -\frac{b^2}{4} \,\omega_{\alpha\beta} \,\omega^{\alpha\beta}\\
s_b^{(2)} &= \left(\frac{1}{2}+\frac{2}{3}\,\text{ln}\ 2 +  \mathcal{C}+\frac{\pi}{6}+\frac{5\,\pi^2}{48}\right)\,\sigma_{\alpha\beta}\, \sigma^{\alpha\beta} -\frac{1}{2}\, \omega_{\alpha\beta} \,\omega^{\alpha\beta}+\frac{1}{6}\,\mathcal{R}\\
\end{split}
\end{equation}
while the vector contribution is given as
\begin{equation}
j_\mu^{(2)}=P_{\mu}^{\nu}\,\left[\frac{3}{4}\,\mathcal{D}_\lambda\sigma_\nu{}^{\lambda}+\frac{1}{2}\,\mathcal{D}_\lambda\omega_\nu{}^{\lambda}\right]=\left(-\frac{3}{4}\,\sigma_{\alpha\beta}\, \sigma^{\alpha\beta}+\frac{1}{2}\,\omega_{\alpha\beta}\, \omega^{\alpha\beta}\right) \,u_\mu  +\frac{3}{4}\,\mathcal{D}_\lambda\sigma_\mu{}^{\lambda}+\frac{1}{2}\,\mathcal{D}_\lambda\omega_\mu{}^{\lambda}
\label{}
\end{equation}	

Now, we use \eqref{simanto}, \eqref{nm2} and \eqref{sg} to write $r_H,n^\mu$ and $\sqrt{g}$  in Weyl covariant form as follows:
\begin{equation}\label{rhcov:eq1}
r_{H} = \frac{1}{b}\left(1+\frac{b^2}{4}\left[\left(\frac{5}{6}+\frac{2}{3}\,\text{ln}\ 2 +  \mathcal{C}+\frac{\pi}{6}+\frac{5\,\pi^2}{48}\right)\, \sigma_{\alpha\beta} \,\sigma^{\alpha\beta} -\frac{1}{2} \,\omega_{\alpha\beta} \,\omega^{\alpha\beta}+\frac{1}{6}\,\mathcal{R} \right]\right)\\
\end{equation}
\begin{eqnarray}\label{rhcov:eq2}
n^\mu &=& \left(1-\frac{b^2}{4}\left[\frac{1}{2}+\text{ln}\ 2 + 3\ \mathcal{C}+\frac{\pi}{4}+\frac{5\,\pi^2}{16}-\left(\frac{3}{2}\,\text{ln}\ 2 +\frac{\pi}{4}\right)^2\right] \,\sigma_{\alpha\beta} \,\sigma^{\alpha\beta}-\frac{b^2}{4} \,\omega_{\alpha\beta} \,\omega^{\alpha\beta}\right) \,u^\mu \nonumber\\
&&\qquad \qquad  +\, b^2 \,P_{\mu}^{\nu}\,\left(\frac{1}{4}\,\mathcal{D}_\lambda\sigma_\nu{}^{\lambda}+\frac{1}{2}\,\mathcal{D}_\lambda\omega_\nu{}^{\lambda}\right) 
\end{eqnarray}
\begin{equation}\label{rhcov:eq3}
\sqrt{g} = \frac{1}{b^3}\,\left(1+\frac{b^2}{4}\left[\left(2+\text{ln}\ 2 +\frac{\pi}{4} \right)\,\sigma_{\alpha\beta}\, \sigma^{\alpha\beta} -\frac{5}{2}\,\omega_{\alpha\beta} \,\omega^{\alpha\beta}+\frac{1}{2}\,\mathcal{R} \right] \ . \right)
\end{equation}

Putting all of these together we can finally obtain the expression for the entropy current: 
\begin{equation}\label{jscov:eq}
\begin{split}
4\,G^{(5)}_N\,b^3 \,J^\mu_S &= \left(1+b^2\,\left[\left(\frac{1}{2}+\frac{1}{4}\text{ln}\ 2 +\frac{\pi}{16} \right] \,\sigma_{\alpha\beta}\, \sigma^{\alpha\beta} -\frac{5}{8}\,\omega_{\alpha\beta} \omega^{\alpha\beta} +\frac{1}{8}\,\mathcal{R} \right]\right) u^\mu \\
&\qquad \qquad +  b^2\, P_{\mu}^{\nu}\,\left(\frac{1}{4}\,\mathcal{D}_\lambda\sigma_\nu{}^{\lambda}+\frac{1}{2}\,\mathcal{D}_\lambda\omega_\nu{}^{\lambda}\right) \\
&= \left(1+b^2\,\left[\left(\frac{1}{4}+\frac{1}{4}\,\text{ln}\ 2 +\frac{\pi}{16} \right)\,\sigma_{\alpha\beta} \,\sigma^{\alpha\beta} -\frac{1}{8}\,\omega_{\alpha\beta} \,\omega^{\alpha\beta} +\frac{1}{8}\,\mathcal{R} \right]\right)\, u^\mu \\
&\qquad \qquad +  b^2 \,\left(\frac{1}{4}\,\mathcal{D}_\lambda\sigma^{\mu\lambda}+\frac{1}{2}\,\mathcal{D}_\lambda\omega^{\mu\lambda}\right) \\
\end{split}
\end{equation}
from which we can read off the coefficients $A$, $B$ and $C$  appearing in the general current \eqref{derivjs:eq}
\begin{equation}\label{actualABC:eq}
\begin{split}
A_1 = \frac{1}{4}+\frac{\pi}{16}+\frac{\text{ln}\ 2}{4}  ;\qquad A_2 &=-\frac{1}{8} ;\qquad A_3 =\frac{1}{8} \\
B_1=\frac{1}{4} ;\qquad& B_2=\frac{1}{2}  \\
C_1 =& \;C_2 = 0 \\
\end{split}
\end{equation}

These coefficients manifestly obey the constraints laid down in \eqref{condnB:eq} and hence, the entropy current derived from gravity obeys the second law. Further, we get the divergence of the entropy current as
\begin{equation}\label{divjs3:eq}
\begin{split}
4\,G^{(5)}_N\,b^3 \,J^\mu_S &= b^2\,\sigma_{\mu\nu}\,\left[\frac{\sigma^{\mu\nu}}{2 \,b} + 2\,\left( \frac{1}{4}+\frac{\pi}{16}+\frac{3}{8}\,\text{ln}\ 2\right)\, u^\lambda\mathcal{D}_\lambda \sigma^{\mu\nu} \right] +\ldots\\
&= \frac{b}{2}\,\left[\sigma^{\mu\nu} + b\ \left( \frac{1}{4}+\frac{\pi}{16}+\frac{3}{8}\,\text{ln}\ 2\right)\,  u^\lambda\mathcal{D}_\lambda \sigma^{\mu\nu} \right]^2 +\ldots\\
\end{split}
\end{equation}
which can alternatively be written in the form
\begin{equation}\label{divjs4:eq}
\begin{split}
T\, \mathcal{D}_\mu J^\mu_S &= 2\, \eta\, \left[\sigma^{\mu\nu} +\frac{\left( \pi+4+6\ \text{ln}\ 2 \right)}{16\pi\,T}\, u^\lambda\mathcal{D}_\lambda \sigma^{\mu\nu} \right]^2 +\ldots\\
\end{split}
\end{equation}
which gives the final expression for the rate of entropy production computed via holography.

\subsection{Ambiguity in the holographic entropy current}
\label{app:ambecur}

We now examine briefly the change in  the coefficients  $A$, $B$ and $C$ parametrizing the arbitrary entropy current,  under the ambiguity shift discussed  in \sec{sec:gravamb}, see  Eq. \eqref{entambig}. In particular, we want to verify explicitly that under such a shift, the entropy production still remains positive semi-definite.

The first kind of ambiguity in the entropy current arises due to the addition of an exact form to the entropy current. The only Weyl covariant exact form that can appear in the entropy current at this order is given by
\begin{equation}
\begin{split}
4 \,G^{(5)}_N \,b^3 \,\delta J^\mu_S  &= \delta\lambda_0 b^2 \mathcal{D}_\nu \omega^{\mu\nu}
\end{split}
\end{equation}
which induces a shift in the above coefficients $B_2\longrightarrow B_2 +\delta\lambda_0$.
 
The second kind shift in the entropy current (due to the arbitrariness in the boundary to horizon map) is parametrised by a vector  $\delta\zeta^\mu$(which is Weyl-invariant) and is given by
\begin{equation}
\begin{split}\label{ambshift:eq}
\delta J^\mu_S  &= \mathcal{L}_{\delta\zeta} J^\mu_S -  J^\nu_S \,\nabla_\nu \delta\zeta^\mu \\
&= \mathcal{D}_\nu\left[J^\mu_S \,\delta\zeta^\nu- J^\nu_S \, \delta\zeta^\mu \right] + \delta\zeta^\mu \, \mathcal{D}_\nu J^\nu_S
\end{split}
\end{equation}
where in the last line we have rewritten the shift in a manifestly Weyl-covariant form.

If we now write down a general derivative expansion for $\delta\zeta^\mu$ as
\begin{equation}
\delta\zeta^\mu = 2\ \delta\lambda_1\ b\ u^\mu + \delta\lambda_2\ b^2\ \l^\mu +\ldots 
\end{equation}
the shift in the entropy current can be calculated using the above identities as 
\begin{equation}\label{ambshift2:eq}
\begin{split}
4 \,G^{(5)}_N \,b^3 \,\delta J^\mu_S  &= \delta\lambda_1\ b^2\ \sigma_{\alpha\beta}\,\sigma^{\alpha\beta}\, u^\mu  +\ldots
\end{split}
\end{equation}
which implies a shift in  the above coefficients given by $A_1 \longrightarrow A_1 + \delta\lambda_1\ $.

Note that both these shifts maintain the constraints listed in \eqref{condnB:eq} and hence, the positive semi-definite nature of the entropy production is unaffected by these ambiguities as advertised.

\section{Wald's ``entropy form''}
\label{wald-app} 

In this section we briefly discuss the notion of a local ``entropy form'',
as defined by Wald,  \cite{Wald:1997qz, Iyer:1994ys,Wald:1993nt}. This is defined using a variational principle for any diffeomorphism invariant Lagrangian ${\mathcal L}$ to derive an expression for the first law of black hole mechanics. We consider a $d+1$ dimensional spacetime with metric $G_{AB}$ which is a solution to ${\mathcal L}$'s equations of motion and denote $\nabla_A$ to be the associated covariant derivative.

\subsection{Stationary black branes}

Let us first consider the case of a stationary black brane,
characterized by a Killing horizon ${\mathcal H}$ which is
generated by a Killing vector $\chi^A$. We normalize $\chi^A$ by the
condition that it satisfies $\chi_A \,\nabla^A \chi^B=\chi^B$ on ${\mathcal H}$ and
assume that ${\mathcal H}$ possesses a bifurcation surface
$\Sigma_b$.

Consider the following $d-1$-form
\begin{align}
{\bf S}_{A_1...A_{d-1}} =- \frac{2\,\pi \sqrt{-G}}{(d-1)!}  \, 
\frac{\partial {\mathcal L}}{\partial
R_{ABCD}} \,\epsilon_{A_1 \cdots A_{d-1}AB} \, \nabla_C \chi_D
\label{entropy-form}
\end{align}
It has been shown in \cite{Iyer:1994ys, Wald:1993nt} that the entropy of the black hole $S$  is then simply the integral of ${\bf S}$ over $\Sigma_b$, and  it satisfies the first law of thermodynamics.  Hence, \eqref{entropy-form} provides a local expression for the entropy-
form. As is to be expected, this expression is not unique, and suffers from ambiguities arising from: (i) the possibility of adding exact derivatives to ${\mathcal L}$, (ii) addition of a $(d-1)$ form to ${\bf S}$ which arises
from the additive ambiguity of the Noether current up to the Hodge dual of an exact
$d$-form, and (iii) the possibility of adding to ${\bf S}$ an exact $d-1$ form without changing the entropy $S$, \cf.  proposition 4.1 of \cite{Iyer:1994ys}.  However, in the discussion that follows, these additional terms will not be important.

It is easy to evaluate the above expression (\ref{entropy-form}) in
case of General Relativity. In this case ${\mathcal
L}=\frac{1}{16\pi \,G^{(d+1)}_N}\, (R+\Lambda)$, and
\begin{align} 
\frac{\partial  {\mathcal L}}{\partial R_{ABCD}}=
\frac{1}{32\pi \,G^{(d+1)}_N}\; (G^{AC}\,G^{BD} - G^{BC}\,G^{AD}).
\label{l-rabcd}
\end{align}
Further, on $\Sigma_b$, 
\begin{align}
\nabla_{[A}\chi_{B]} = {\bf n}_{AB}
\label{pre-binormal}
\end{align}
where ${\bf n}_{AB}$ is the binormal to $\Sigma_b$, defined by
\begin{align}
{\bf
n}_{AB}= N_A \chi_B - N_B \chi_A,
\label{binormal} 
\end{align}
where $N^A$ is the ``ingoing''
future-directed null vector, normalized such that $N^A\,\chi_A=-1$
\footnote{This normalization, together with fact that $N^A, \chi^A$
are both normal to $\Sigma_b$ uniquely fixes $N^A$.}. It is easy to
show that the volume element on $\Sigma_b$ is given by (see
Eq. (12.5.34) of \cite{Wald:1984gr}):
\begin{align}
\Omega_{A_1 ... A_{d-1}} = -\frac{\sqrt{-G}}{2(d-1)!}\, 
\epsilon_{A_1...A_{d-1} A B} \; {\bf n}^{AB}
=- \frac{\sqrt{-G}}{(d-1)!} \,
\epsilon_{A_1...A_{d-1} A B} \;\nabla_C\chi_D  \;G^{AC}\,G^{BD}
\label{volume}
\end{align}
Putting all this together, the entropy form ${\bf S}$ becomes
\begin{align}
{\bf S}_{A_1...A_{d-1}}= 
\frac1{4\, G^{(d+1)}_N}\, \Omega_{A_1 ... A_{d-1}}
\label{GR-q}
\end{align}
The total entropy is given by the integral
\begin{align} 
S= &\int_{\Sigma_b} {\bf S}_{A_1 ... A_{d-1}}
dx^{A_1} \wedge ... \wedge dx^{A_{d-1}}  
=\frac{1}{4\,G^{(d+1)}_N}\,  {\rm Area}\left(\Sigma_b\right)
\end{align} 
thus reproducing the usual Bekenstein-Hawking formula. For stationary black holes the area of the bifurcation surface of course coincides with the area of the black hole horizon.

To be explicit, let us consider the example of the five dimensional stationary  (boosted) black brane solution  which is given by \eqref{formmet}, with $\epsilon = 0$ and $b$ and $u^\mu$ constants (independent of $x^\mu$). 
The horizon is located at $r=r_H \equiv 1/b$.  Let us consider a spacelike slice $\Sigma \subset {\mathcal H}$ defined by $u_\mu \,dx^\mu=0$. The binormal ${\bf n}_{AB}$ to this surface
\eqref{binormal} is given in terms of the null vectors$\chi^A$ and $N^A$. We have the 
normalized Killing vector  $\chi^A \frac{\partial}{\partial X^A}  =\frac{1}{\kappa} 
u^\mu \frac{\partial}{\partial x^\mu}$ and  $N_A dX^A = \kappa\,(2 \, dr - r^2 \, f(br)\, u_\mu dx^\mu)$.   Here $\kappa=\frac{1}{2}\, \left(r^2\, \partial_r f(br)\right)|_{r=r_H}$.  

From \eqref{binormal} we find that the only
non-vanishing components of the entropy $(d-1$)-form ${\bf S}$ are
given by
\begin{align}
{\bf S}_{\mu_1 \mu_2 ...\mu_{d-1}} &= 
\frac{r_H^{d-1}}{4\,G^{(d+1)}_N\, (d-1)!} \, u^\mu\,
\epsilon_{\mu \mu_1 \mu_2... \mu_{d-1}} 
= \frac{\sqrt{h}}{4\,G^{(d+1)}_N\, (d-1)!} \,
\epsilon_{\mu \mu_1 \mu_2... \mu_{d-1}} \,\frac{u^\mu}{u^v} \ . 
\label{wald-static-area}
\end{align} 
We have used the fact that on the $(d-1)$-surface $\Sigma$, 
$\sqrt{h} = r_H^{d-1}\, u^v$.

The entropy form $a$, given in (\ref{enttf4}),  agrees with the above expression:
\begin{align}
a = {\bf S}_{\mu_1 \mu_2... \mu_{d-1}} dx^{\mu_1}\wedge dx^{\mu_2}
\wedge...\wedge dx^{\mu_{d-1}}
\end{align}
where we note that the vector $n^\mu$ in (\ref{enttf4}) becomes
equal to $u^\mu$ in the static case (we will discuss the dynamical situation below).

It is interesting to note that in case of higher
derivative gravity, the entropy form has terms in addition
to the area-form. For example, 
in case of Lovelock gravity, with Lagrangian density
\begin{align} 
{\mathcal L}=\frac{1}{16\pi\,G^{(d+1)}_N} \,R+\alpha\, \left( R_{ABCD}\,R^{ABCD} 
-4\, R_{AB} \,R^{AB} + R^2\right),   
\end{align} 
the entropy form ${\bf S}$ as defined by \eqref{entropy-form} (see  Eq. (72) of 
\cite{Iyer:1994ys}) leads to an entropy, which has contributions from the first Chern class of the bifurcation surface:
\begin{align} 
S=\frac{1}{4\,G^{(d+1)}_N} \,{\rm Area}\left(\Sigma_b\right) +8\pi\, \alpha\, \int_{\Sigma_b} R^{(d-1)}\,  
\sqrt{g^{(d-1)}}\, d^{d-1}x.
\end{align}

\subsection{Dynamical horizons}

The horizon of dynamical black holes (such as the generic situation with \eqref{formmet}) is not generated by a Killing field and generically one doesn't have a bifurcation surface. So the formula (\ref{entropy-form})
cannot be applied as such. However, as argued in
\cite{Iyer:1994ys}, the simplest way to proceed in this case is to develop a notion of a local bifurcation surface, and construct a  ``local Killing field'' $\chi$. In case of General Relativity, this leads to a definition of the entropy $(d-1)$ form ${\bf S}$ as in 
(\ref{entropy-form}); the main distinction is that $\nabla_C \chi_D$ is interpreted as the binormal ${\bf n}_{CD}$ only in a sufficiently small neighbourhood close to the initially chosen surface.  Hence locally, we can continue as before using the result (\ref{volume}) to arrive at
the expression
\begin{align}
{\bf S}_{\mu_1 \mu_2 \ldots \mu_{d-1}} &
= \frac{\sqrt{h}}{4\, G^{(d+1)}_N} \;  
\epsilon_{\mu \mu_1 \mu_2 \ldots \mu_{d-1}} \,\frac{n^\mu}{n^v} 
\label{wald-dyn-area}
\end{align}
Here in constructing the binormal (\ref{binormal}) we have used the
fact that $\chi^A \propto n^A $ on ${\mathcal H}$ where $n^A$ is
defined by (\ref{nxidef}). As explained in \cite{Iyer:1994ys}, at any
point $p$ on a spacelike surface $\sigma \subset {\mathcal H}$ we can choose coordinates
such that the expression (\ref{entropy-form}) for the entropy remains
correct (in particular, one can choose to ensure that the additional terms arising from ambiguities in defining ${\bf S}$ vanish), so that the above derivation of (\ref{wald-dyn-area}) remains valid. It is easy to see that the entropy form $a$, given in (\ref{enttf4}),
agrees with the above expression in the dynamical case as well.

\subsection{Second law}

We saw above, in case of General Relativity in arbitrary dimensions,
that the Wald definition of entropy leads to the area-form on the
horizon. The divergence of the entropy current therefore is
nonnegative as a consequence of Hawking's area theorem and hence obeys the
second law of thermodynamics (assuming cosmic censorship). Recall that area theorem requires that the energy conditions hold; physically, the only when gravity is attractive are we guaranteed area increase. However, in case of higher derivative theories, it is not clear whether the second law is obeyed \cite{Wald:1993nt,Iyer:1994ys} by the Wald entropy (\cf., \cite{Jacobson:1993vj,Jacobson:1994qe} for a discussion in certain special classes of higher derivative theories). This is simply because higher derivative theories violate the energy conditions and the situation is further complicated by the fact that the entropy starts to depend on the intrinsic geometry of the black hole horizon.
On the other hand, from the viewpoint of the boundary theory, $\alpha'$ corrections
simply provide a one-parameter deformation of various parameters of
the fluid which must continue to obey the second law of thermodynamics. It would be interesting to resolve this puzzle (see \sec{sec:discuss} for comments).

\section{Independent data in fields up to third order}
\label{app:inddata}

There are 16, 40 and 80 independent components at first, second and third 
orders in the Taylor expansion of velocity and temperature.\footnote{For each independent function we count the number of independent partial derivatives at a given order; for the temperature we have $\partial_\mu T$, $\partial_\mu \partial_\nu T$, \etc.} These pieces of  data are not all independent; they are constrained by equations of motion.  The relevant equations of 
motion are the conservation of the stress tensor and its first and second 
derivatives\footnote{The relevant equations are just the moments of the conservation equation which arise as local constraints at higher orders.} (at our spacetime point) which are $4, 16$ and $40$ respectively 
in number.\footnote{As $T^{\mu\nu}$ is not homogeneous in the derivative 
expansion, these equations of motion mix terms of different order in this 
expansion.} The terms that appear in the three kinds of equations listed above 
start at first, second and third order respectively. Consequently these 
equations may be used to cut down the independent data in Taylor series 
coefficients of the velocity and temperature at first second 
and third order to 12, 24 and 40 components respectively. We will now 
redo this counting keeping track of the $SO(3)$ transformation properties of 
all fields. 

Let us list degrees of freedom by the vector $(a, b, c, d, e)$ where $a$ 
represents the number of $SO(3)$ scalars ({\bf 1}), $b$ the number of $SO(3)$  vectors ({\bf 3}), \etc. Working up to third order we encounter terms transforming in at most the ${\bf 9}$ representation of $SO(3)$. In this notation, the number of degrees of 
freedom in Taylor coefficients are  $(2,3,1,0,0)$, $(3,5,3,1,0)$, and 
$(4,7, 5, 3, 1)$ at first, second and third order respectively. 
The number of equations of motion are $(1,1,0,0,0)$, $(2,3,1,0,0)$ and $(3, 5, 3,1,0)$ 
respectively (note that the number of equations of motion at order $n+1$ is 
the same as the number of variables at order $n$). It follows from 
subtraction that the number of unconstrained variables at zeroth, first, second 
and third order respectively can be chosen to be $(1,1,0,0,0)$, $(1,2,1,0,0)$, 
$(1,2,2,1,0)$ and $(1,2,2,1,1)$. This choice is convenient in checking the statements about the non-negativity of the divergence of the entropy current at third order explicitly.

\bibliographystyle{utphys}

\providecommand{\href}[2]{#2}\begingroup\raggedright\endgroup

\end{document}